\documentclass{article}
\usepackage{graphicx} 
\usepackage{amsmath}
\usepackage{amssymb}
\usepackage{mathtools}
\usepackage{placeins}
\usepackage{physics}
\usepackage{xcolor}
\usepackage{dcolumn}
\usepackage{bm}
\usepackage{caption}
\usepackage{url}
\usepackage{amsthm}
\usepackage{authblk}
\usepackage{placeins}
\usepackage{float}

\author[1]{L. Yıldız\thanks{\texttt{li.yildiz.na@gmail.com}}}
\author[2]{D. Kaykı\thanks{\texttt{deha.kayki@ogr.iu.edu.tr}}}
\author[3]{E. Güdekli\thanks{\texttt{gudekli@istanbul.edu.tr}}}

\affil[1,2,3]{Department of Physics, Faculty of Science, Istanbul University, Istanbul 34134, Turkey}

\title{Linear Perturbations and Multi-Probe Diagnostics in Dark-Sector Selective $f(R,T_{\chi})$ Gravity}

\date{}

\begin{document}

\maketitle
\begin{abstract}
We develop a dark-sector selective trace-coupled extension of gravity in which the matter--curvature coupling depends exclusively on the trace of the dark-matter energy--momentum tensor, $T_{\chi}$, defined from a canonical dark-matter field $\chi$. This construction provides a microphysically specified trace sector, removes the usual matter-Lagrangian ambiguity of $f(R,T)$-type models, and preserves minimal coupling of visible matter by design. We derive the full field equations, the exact dark-sector exchange structure, and the linear scalar-perturbation system in gauge-ready form. In the sub-horizon regime, we derive effective modified-gravity functions governing structure growth and light deflection, and show that the model generically produces correlated, scale- and time-dependent departures from General Relativity in growth and lensing observables. Building on this structure, we formulate a perturbation-focused multi-probe framework based on redshift-space distortions, weak lensing, and CMB lensing, explicitly targeting degeneracy breaking beyond background-expansion tests. The analysis establishes the action-level and perturbation-level foundations of the model and provides a conservative, reproducible framework for translated linear-regime constraints within a dark-sector selective modified-gravity setting.
\end{abstract}


\noindent\textbf{Highlights}\\
\begin{itemize}
  \item Dark-sector-only $T_{\chi}$ coupling removes direct visible-matter trace coupling at the action level
  \item Microphysical canonical-$\chi$ construction removes the $L_m$ ambiguity of $f(R,T)$-type models
  \item Linear scalar perturbations yield scale- and time-dependent growth and lensing signatures
  \item Perturbation-level multi-probe diagnostics target background-only degeneracies in modified gravity
\end{itemize}

\section{Introduction}
\label{sec:introduction}

The physical nature of the dark sector remains one of the most persistent open problems in
fundamental physics and cosmology \cite{HutererShafer2018DarkEnergyReview,Amendola2018EuclidFundamentalPhysics}. The concordance $\Lambda$CDM framework provides an
excellent effective description of a broad range of observations\cite{Planck2018Parameters}, yet it leaves the
microphysics of dark matter and dark energy unspecified \cite{HutererShafer2018DarkEnergyReview,Amendola2018EuclidFundamentalPhysics}. This motivates theoretically
controlled extensions that are both internally consistent and empirically testable \cite{Ishak2019TestingGRCosmology,Amendola2018EuclidFundamentalPhysics}.
In particular, many beyond-$\Lambda$CDM scenarios are nearly degenerate at the level of
background expansion, so their viability must be assessed primarily through
\emph{perturbation-sensitive} signatures such as the growth of cosmic structure and the
deflection of light \cite{Linder2005GrowthExpansion,Ishak2019TestingGRCosmology,BoubekeurGiusarmaMenaRamirez2014MGStatus}. A modified-gravity framework aimed at the dark sector should therefore
deliver predictive, falsifiable correlations between growth and lensing observables rather
than relying on background fits alone \cite{Ishak2019TestingGRCosmology,Amendola2018EuclidFundamentalPhysics}.

Trace-coupled modifications of gravity provide a phenomenologically rich arena to explore
such signatures \cite{Harko2011fRT,Koyama2016CosmoTestsMG}. In the standard $f(R,T)$ class, the gravitational Lagrangian depends on the
Ricci scalar $R$ and the trace of the matter stress-energy tensor $T$, enabling an
effective interaction between matter and geometry \cite{Harko2011fRT}.
However, two recurrent issues arise in generic $f(R,T)$ constructions \cite{Harko2011fRT,Carvalho2021LmAmbiguityfRT}.
First, the theory can inherit an ambiguity associated with the choice of the matter
Lagrangian density $L_m$ and the related tensor $\Theta_{\mu\nu}$, which affects the precise
form of the field equations and the exchange terms. Second, if visible matter couples
directly to the modified sector, baryonic fifth-force constraints and local tests of gravity
can severely restrict the parameter space \cite{Joyce2015BeyondCSM,Koyama2016CosmoTestsMG}. These issues become particularly acute when
one attempts to build a predictive perturbation theory suitable for confrontation with
large-scale-structure and lensing measurements \cite{Ishak2019TestingGRCosmology,Koyama2016CosmoTestsMG}.

In this work we address both problems by introducing a \emph{dark-sector selective}
trace coupling in which the modified-gravity dependence on the matter trace is restricted
to the \emph{dark-matter} sector only, $T \rightarrow T_{\chi}$.
Visible matter components (baryons and radiation) remain minimally coupled by construction,
suppressing direct baryonic fifth-force effects at the level of the fundamental action.
In addition, we define the dark matter microphysically by a canonical field $\chi$,
so that $T_{\chi}$ and $\Theta^{(\chi)}_{\mu\nu}$ are uniquely determined.
This choice removes the standard $L_m$ ambiguity and yields an unambiguous exchange structure
between the dark sector and the modified gravitational dynamics.

The construction is intended as a minimal, physically interpretable, and observationally
testable effective framework, rather than a unique UV-complete theory. Its purpose is to
isolate curvature--dark-sector coupling effects in the sector where late-time cosmological
degeneracies are most likely to appear, while preserving a clean separation from visible-sector
dynamics at the level of the fundamental action. This separation does not imply that
visible-sector observables are unaffected; rather, their response is mediated indirectly through
the modified gravitational dynamics, not through a direct trace coupling in the matter action.
In this sense, the model provides a controlled setting in which the impact of selective trace
coupling can be tracked consistently from background evolution to linear perturbations and,
ultimately, to multi-probe observables.

Our primary goal is to develop a perturbation-complete and observationally relevant
framework for this class of models.
We derive the full field equations and the exact dark-sector exchange relations implied by
the selective coupling, and we obtain the linear scalar perturbation system in a gauge-ready
form suitable for direct implementation in numerical pipelines.
In the sub-horizon regime, we show how the theory can be mapped to effective modified-gravity
functions that control the clustering of matter and the deflection of light, enabling
predictions for redshift-space distortions (via $f\sigma_8$) and for weak and CMB lensing \cite{PogosianSilvestri2016SigmaMu,Ishak2019TestingGRCosmology,Koyama2016CosmoTestsMG}.
This multi-probe focus is essential to break background-level degeneracies and to isolate
signatures that distinguish the model from General Relativity and $\Lambda$CDM \cite{Linder2005GrowthExpansion,Ishak2019TestingGRCosmology,Amendola2018EuclidFundamentalPhysics}.

\paragraph{Positioning relative to other modified-gravity frameworks.}
The construction is intended to occupy an intermediate position between fully generic
trace-coupled models and purely phenomenological late-time parameterizations.
Compared with generic $f(R,T)$ theories, the restriction $T\to T_\chi$ and the microphysical
canonical-scalar realization of the dark sector remove the usual matter-Lagrangian ambiguity
and prevent direct trace coupling to visible matter at the level of the fundamental action,
while retaining a non-trivial matter--curvature exchange structure in the dark sector
\cite{Harko2011fRT}. Compared with pure $f(R)$ models, the selective $R\,T_\chi$ and
$T_\chi^2$ channels enlarge the perturbation-level phenomenology by introducing an explicit
dark-sector trace response in addition to the standard curvature (scalaron) sector
\cite{DeFeliceTsujikawa2010fRReview}. Finally, unlike model-independent effective descriptions
that parameterize modified growth and lensing directly at the level of linear perturbations
(e.g.\ EFT/Horndeski-inspired or $(\mu,\Sigma)$-type approaches), the framework retains
an explicit action-level parametrization with a transparent map from Lagrangian coefficients
$(\alpha,\xi,\beta)$ to background and perturbation quantities, which is useful for
implementation-ready and falsifiable multi-probe inference analyses
\cite{Gubitosi2013EFTDE,BelliniSawicki2014Alpha,Ishak2019GRCosmoReview}.

From the viewpoint of local-gravity phenomenology, the selective restriction $T\to T_\chi$
is also advantageous: because visible matter remains minimally coupled in the matter action,
the model avoids direct baryonic trace-coupling effects at the fundamental level. This does
not eliminate the need for screening or local-gravity consistency tests, since visible-sector
observables can still be affected indirectly through the modified gravitational dynamics, but
it provides a cleaner starting point than generic trace-coupled constructions for incorporating environment-dependent screening mechanisms.

The paper is organized as follows.
In Sec.~\ref{sec:framework} we introduce the dark-sector selective $f(R,T_{\chi})$ action,
define the microphysical dark-matter sector, and derive the field equations together with
their conservation and exchange structure.
In Sec.~\ref{sec:background} we specialize the theory to an FLRW background and obtain the
modified cosmological evolution equations.
In Sec.~\ref{sec:perturbations} we derive the linear perturbation equations and identify the
effective functions governing growth and lensing.
In Sec.~\ref{sec:obsfe_inrence} we connect the theoretical predictions to perturbation-sensitive
observables and outline a multi-probe inference strategy.
We conclude in Sec.~\ref{sec:conclusions} with a summary and outlook.

\section{Dark-sector selective $f(R,T_{\chi})$ framework}
\label{sec:framework}

\subsection{Conventions and definitions}
We adopt metric signature $(-,+,+,+)$ and units $c=\hbar=1$.
The covariant derivative is $\nabla_{\mu}$ and $\Box \equiv g^{\mu\nu}\nabla_{\mu}\nabla_{\nu}$.
We define the partial derivatives of the gravitational function as
\begin{equation}
f_{R} \equiv \frac{\partial f}{\partial R},
\qquad
f_{T} \equiv \frac{\partial f}{\partial T_{\chi}}.
\label{eq:fR_fT_defs}
\end{equation}

\subsection{Action and matter content}
We consider a trace-coupled extension in which the \emph{gravitational trace dependence} is restricted to the dark-matter trace $T_{\chi}$, while visible components remain minimally coupled.
The action is
\begin{equation}
S=\int \mathrm{d}^4x\,\sqrt{-g}\left[
\frac{1}{2\kappa^2}\, f(R,T_{\chi})
+\mathcal{L}_b+\mathcal{L}_{\mathrm{rad}}+\mathcal{L}_{\chi}
\right],
\label{eq:action_main}
\end{equation}
where $\kappa^2 \equiv 8\pi G$ and $\mathcal{L}_b$ and $\mathcal{L}_{\mathrm{rad}}$
denote baryons and radiation.

This construction is adopted as a dark-sector selective extension of the $f(R,T)$ framework \cite{Harko2011fRT},
in which the trace dependence of the gravitational sector is restricted to $T_\chi$ rather than
the total matter trace. The motivation is twofold: it preserves a clean phenomenological
separation between visible-sector dynamics and dark-sector backreaction channels, and it isolates
the modified coupling to the sector where late-time cosmological degeneracies and beyond-$\Lambda$CDM
signatures are most naturally probed in a controlled way. In this sense, the model is treated throughout as a minimal, physically interpretable,
and observationally testable effective framework, rather than a unique UV-complete theory.

To remove the usual matter-Lagrangian ambiguity present in generic $f(R,T)$ scenarios,
we define dark matter microphysically as a canonical scalar field $\chi$,
\begin{equation}
\mathcal{L}_{\chi}=-\frac{1}{2}\,g^{\mu\nu}\partial_{\mu}\chi\,\partial_{\nu}\chi - V(\chi).
\label{eq:Lchi}
\end{equation}
The corresponding stress-energy tensor follows from
$T^{(\chi)}_{\mu\nu} \equiv -\frac{2}{\sqrt{-g}}
\frac{\delta(\sqrt{-g}\mathcal{L}_{\chi})}{\delta g^{\mu\nu}}$, giving
\begin{equation}
T^{(\chi)}_{\mu\nu}
=\partial_{\mu}\chi\,\partial_{\nu}\chi
-g_{\mu\nu}\left[\frac{1}{2}(\partial\chi)^2+V(\chi)\right],
\label{eq:Tmn_chi}
\end{equation}
with $(\partial\chi)^2 \equiv g^{\rho\sigma}\partial_{\rho}\chi\,\partial_{\sigma}\chi$.
Its trace is
\begin{equation}
T_{\chi}\equiv g^{\mu\nu}T^{(\chi)}_{\mu\nu}
=-(\partial\chi)^2-4V(\chi).
\label{eq:Tchi_def}
\end{equation}

With the canonical choice \eqref{eq:Lchi}, both $T_\chi$ and the metric variation entering
$\Theta^{(\chi)}_{\mu\nu}$ are fixed uniquely, so the $f_T$-induced exchange terms are defined
without the matter-Lagrangian ambiguity that affects generic phenomenological $f(R,T)$ implementations \cite{Harko2011fRT,Carvalho2021LmAmbiguityfRT}.

The restriction $T\rightarrow T_{\chi}$ in Eq.~\eqref{eq:action_main} defines a
\emph{dark-sector selective} matter--curvature coupling. Visible matter is assumed to be
minimally coupled through $\mathcal{L}_b$ and $\mathcal{L}_{\mathrm{rad}}$, so that standard
local constraints associated with direct baryonic couplings are avoided at the level of the
fundamental action \cite{Joyce2015BeyondCSM,Koyama2016CosmoTestsMG}. This does not imply that visible-sector observables are unaffected; rather,
their response is mediated indirectly through the modified gravitational dynamics, not through a
direct trace coupling in the matter action. In the next subsection we derive the modified field equations and the exact exchange structure induced by $f_T\neq 0$, which will be the basis for
the background and linear perturbation analysis.

\subsection{Field equations and dark-sector exchange structure}
\label{subsec:field_equations}

Varying the action \eqref{eq:action_main} with respect to the metric yields the modified
field equations. Writing $f_R \equiv \partial f/\partial R$ and $f_T \equiv \partial f/\partial T_\chi$,
one obtains
\begin{align}
f_R R_{\mu\nu} - \frac{1}{2} f\, g_{\mu\nu}
+ \left(g_{\mu\nu}\Box - \nabla_{\mu}\nabla_{\nu}\right) f_R
&=
\kappa^2\left(T^{(b)}_{\mu\nu}+T^{(\mathrm{rad})}_{\mu\nu}+T^{(\chi)}_{\mu\nu}\right)
- f_T\left(T^{(\chi)}_{\mu\nu}+\Theta^{(\chi)}_{\mu\nu}\right),
\label{eq:field_eqs_general}
\end{align}
where the tensor $\Theta^{(\chi)}_{\mu\nu}$ is defined by
\begin{equation}
\Theta^{(\chi)}_{\mu\nu}
\equiv g^{\alpha\beta}\frac{\delta T^{(\chi)}_{\alpha\beta}}{\delta g^{\mu\nu}}.
\label{eq:Theta_def}
\end{equation}
Because the trace coupling is restricted to $T_\chi$, only the dark-matter sector enters
the $f_T$ term; visible matter remains minimally coupled.

\subsubsection*{Closed form of $\Theta^{(\chi)}_{\mu\nu}$ for canonical $\chi$}
For the canonical scalar field defined in Eq.~\eqref{eq:Lchi}, $\Theta^{(\chi)}_{\mu\nu}$
is uniquely determined. Using the standard identity for minimally coupled matter,
one finds the closed form \cite{Harko2011fRT,Carvalho2021LmAmbiguityfRT}
\begin{equation}
\Theta^{(\chi)}_{\mu\nu}
= -2\,T^{(\chi)}_{\mu\nu} + g_{\mu\nu}\mathcal{L}_{\chi}
= g_{\mu\nu}\left[\frac{1}{2}(\partial\chi)^2+V(\chi)\right]
-2\,\partial_{\mu}\chi\,\partial_{\nu}\chi.
\label{eq:Theta_chi_closed}
\end{equation}
A particularly useful simplification is
\begin{equation}
T^{(\chi)}_{\mu\nu}+\Theta^{(\chi)}_{\mu\nu}
= -\,\partial_{\mu}\chi\,\partial_{\nu}\chi,
\label{eq:TplusTheta_simple}
\end{equation}
which makes the structure of the trace coupling especially transparent.

The canonical scalar-field description \eqref{eq:Lchi} is adopted as a minimal microphysical
realisation of the dark sector that fixes $T_\chi$ and the metric variation entering
$\Theta_{\mu\nu}^{(\chi)}$ unambiguously, thereby enabling a perturbation-complete and
implementation-ready formulation of the theory. We do not interpret $\chi$ as a unique
fundamental particle-physics model of dark matter; rather, it serves as a representative
low-energy effective degree of freedom capturing a broad class of scalar dark-matter scenarios
(e.g.\ ultralight/axion-like fields or moduli in cosmology) at the level relevant for background
and linear-perturbation dynamics \cite{Marsh2016AxionCosmology,Hui2017UltralightScalarsDM,Kane2015CosmologicalModuliReview}. The framework can be straightforwardly generalised to other
dark-matter Lagrangians if desired, with the present choice providing the minimal benchmark
required for transparent multi-probe predictions. Extensions to non-canonical scalar sectors or fluid/field effective descriptions can be treated within the same selective-coupling strategy, but are beyond the scope of the present perturbation-benchmark study.

\subsubsection*{Conservation laws and dark-sector exchange}
Since baryons and radiation appear only through minimal couplings in
$\mathcal{L}_b$ and $\mathcal{L}_{\mathrm{rad}}$, they satisfy the standard conservation laws,
\begin{equation}
\nabla^{\mu}T^{(b)}_{\mu\nu}=0,
\qquad
\nabla^{\mu}T^{(\mathrm{rad})}_{\mu\nu}=0.
\label{eq:visible_conservation}
\end{equation}
In contrast, the dark sector generally exchanges energy-momentum with the modified gravity
sector when $f_T\neq 0$. The non-conservation equation implied by the Bianchi identity and
Eq.~\eqref{eq:field_eqs_general} can be written as \cite{Harko2011fRT}
\begin{equation}
\nabla^{\mu}T^{(\chi)}_{\mu\nu}
=
\frac{f_T}{\kappa^2-f_T}
\left[
\left(T^{(\chi)}_{\mu\nu}+\Theta^{(\chi)}_{\mu\nu}\right)\frac{\nabla^{\mu} f_T}{f_T}
+\nabla^{\mu}\Theta^{(\chi)}_{\mu\nu}
-\frac{1}{2}\nabla_{\nu}T_{\chi}
\right].
\label{eq:chi_nonconservation}
\end{equation}
Using Eq.~\eqref{eq:TplusTheta_simple}, the first term becomes explicitly proportional to
$\partial_{\mu}\chi\,\partial_{\nu}\chi$, making the \emph{dark-sector selective} nature of
the exchange manifest.

\section{Background cosmology}
\label{sec:background}

\subsection{FLRW setup}
We consider a spatially flat FLRW background with cosmic time $t$,
\begin{equation}
\mathrm{d}s^2 = -\mathrm{d}t^2 + a^2(t)\,\delta_{ij}\mathrm{d}x^i \mathrm{d}x^j,
\label{eq:FLRW_metric}
\end{equation}
where $a(t)$ is the scale factor and $H\equiv \dot a/a$ is the Hubble rate.
For this background,
\begin{equation}
R = 6\left(2H^2+\dot H\right),
\label{eq:Ricci_FLRW}
\end{equation}
and we assume a homogeneous dark-matter field $\chi=\chi(t)$.

The standard background densities of baryons and radiation are denoted by $\rho_b$ and
$\rho_{\rm rad}$, with pressures $p_b\simeq 0$ and $p_{\rm rad}=\rho_{\rm rad}/3$.
For the canonical field $\chi$, the background energy density and pressure are 
\begin{equation}
\rho_{\chi}=\frac{1}{2}\dot\chi^2+V(\chi),
\qquad
p_{\chi}=\frac{1}{2}\dot\chi^2-V(\chi),
\label{eq:rho_p_chi}
\end{equation}
so that $\rho_{\chi}+p_{\chi}=\dot\chi^2$.
The trace defined in Eq.~\eqref{eq:Tchi_def} becomes
\begin{equation}
T_{\chi} = \dot\chi^2 - 4V(\chi),
\label{eq:Tchi_background}
\end{equation}
and hence $f_R(t)$ and $f_T(t)$ are time dependent through $R(t)$ and $T_{\chi}(t)$.

\subsection{Modified Friedmann equations}
\label{subsec:modified_friedmann}

Using Eq.~\eqref{eq:field_eqs_general} and the identity
$T^{(\chi)}_{\mu\nu}+\Theta^{(\chi)}_{\mu\nu}=-\partial_{\mu}\chi\,\partial_{\nu}\chi$
from Eq.~\eqref{eq:TplusTheta_simple}, the $00$ component yields
\begin{equation}
3H^2 f_R
=
\kappa^2\left(\rho_b+\rho_{\rm rad}+\rho_{\chi}\right)
+ f_T\,\dot\chi^2
+\frac{1}{2}\left(f_R R - f\right)
-3H\dot f_R .
\label{eq:Friedmann_00}
\end{equation}
The spatial components give
\begin{equation}
-\left(2\dot H+3H^2\right) f_R
=
\kappa^2\left(p_{\rm rad}+p_{\chi}\right)
-\frac{1}{2}\left(f_R R - f\right)
+\ddot f_R + 2H\dot f_R .
\label{eq:Friedmann_ii}
\end{equation}
Combining Eqs.~\eqref{eq:Friedmann_00} and \eqref{eq:Friedmann_ii} produces the useful form
\begin{equation}
-2\dot H\, f_R
=
\kappa^2\left[\left(\rho_b+p_b\right)+\left(\rho_{\rm rad}+p_{\rm rad}\right)+\left(\rho_{\chi}+p_{\chi}\right)\right]
+ f_T\,\dot\chi^2
+\ddot f_R - H\dot f_R ,
\label{eq:Raychaudhuri}
\end{equation}
where $p_b\simeq 0$.

\begin{figure}
    \centering
    \includegraphics[width=0.85\linewidth]{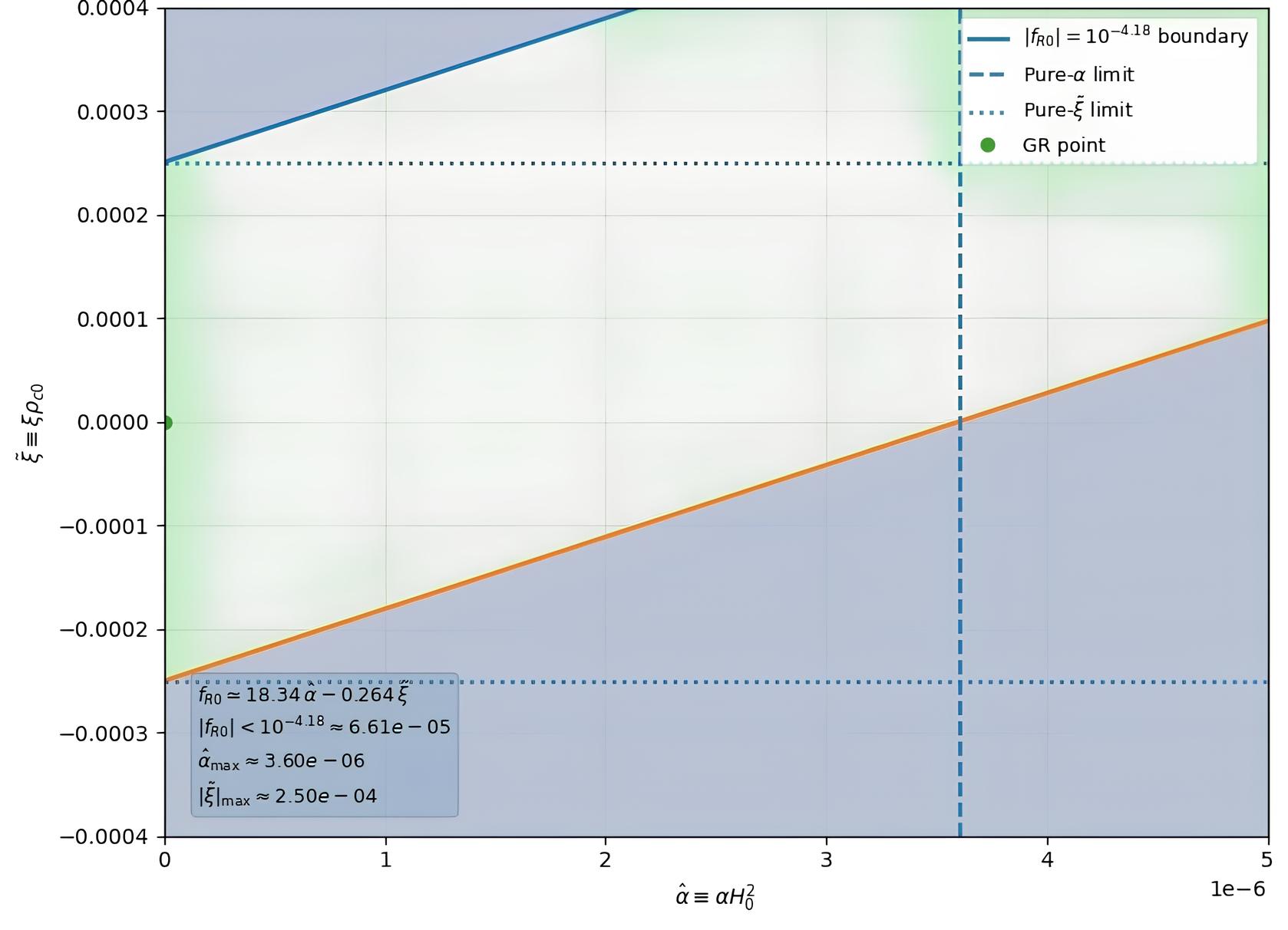}
    \caption{
Translated late-time constraint in the $(\hat{\alpha},\tilde{\xi})$ plane for the polynomial model
$f(R,T_\chi)=R+\alpha R^2+\xi R T_\chi+\beta T_\chi^2$, using the benchmark cap
$\log_{10}|f_{R0}|<-4.18$ and the dust-level approximation $T_{\chi0}\simeq-\rho_{\chi0}$.
Here $\hat{\alpha}\equiv \alpha H_0^2$ and $\tilde{\xi}\equiv \xi\rho_{c0}$.
The shaded strip shows the allowed region satisfying
$|f_{R0}|=|2\alpha R_0+\xi T_{\chi0}|<10^{-4.18}$, illustrating the parameter-combination
degeneracy beyond the pure-$\alpha$ and pure-$\tilde{\xi}$ limits (dashed/dotted lines).
}
    \label{fig:1}
\end{figure}

Figure~\ref{fig:1} makes explicit that the late-time bound on $f_{R0}$ constrains primarily
the combination $2\alpha R_0+\xi T_{\chi0}$, leading to a degeneracy direction in the
$(\hat\alpha,\tilde\xi)$ plane. In the remainder of the paper we therefore treat the model
parameters in a genuinely \emph{multi-probe} way (and, in practice, through normalized parameter
combinations): background information alone constrains only limited combinations, whereas growth
and lensing respond differently to the scalaron-mediated (curvature) sector and to the selective
trace source \cite{Linder2005GrowthExpansion,Ishak2019TestingGRCosmology,Koyama2016CosmoTestsMG,PogosianSilvestri2016SigmaMu}. Combining $f\sigma_8$ with weak/CMB lensing therefore provides the main route to
break this degeneracy at the level of linear perturbations \cite{Ishak2019TestingGRCosmology,Amendola2018EuclidFundamentalPhysics,PogosianSilvestri2016SigmaMu}..

\subsection{Visible-sector conservation and dark-sector evolution}
\label{subsec:background_conservation}

Because baryons and radiation are minimally coupled, Eq.~\eqref{eq:visible_conservation}
implies the standard continuity equations,
\begin{equation}
\dot\rho_b + 3H\rho_b = 0,
\qquad
\dot\rho_{\rm rad} + 4H\rho_{\rm rad} = 0.
\label{eq:rho_b_rad}
\end{equation}

The dark sector generally exchanges energy with the modified gravity sector when $f_T\neq 0$  \cite{Harko2011fRT}.
A convenient way to encode the background dynamics is through the modified equation of motion
obtained by varying the action with respect to $\chi$. Denoting
$V_{,\chi}\equiv \mathrm{d}V/\mathrm{d}\chi$, the homogeneous field satisfies
\begin{equation}
\left(1+\frac{f_T}{\kappa^2}\right)\left(\ddot\chi+3H\dot\chi\right)
+\frac{\dot f_T}{\kappa^2}\,\dot\chi
+\left(1+\frac{2f_T}{\kappa^2}\right)V_{,\chi}
=0.
\label{eq:chi_background_eom}
\end{equation}
Equivalently, the dark-matter energy density obeys a modified continuity equation,
\begin{equation}
\dot\rho_{\chi}+3H\left(\rho_{\chi}+p_{\chi}\right)=Q_{\chi},
\label{eq:chi_continuity_Q}
\end{equation}
where, using Eq.~\eqref{eq:chi_background_eom},
\begin{equation}
Q_{\chi}
=
\dot\chi\left(\ddot\chi+3H\dot\chi+V_{,\chi}\right)
=
-\frac{\dot f_T}{\kappa^2+f_T}\,\dot\chi^2
-\frac{f_T}{\kappa^2+f_T}\,V_{,\chi}\dot\chi .
\label{eq:Qchi_explicit}
\end{equation}
This term quantifies the background-level energy exchange induced by the selective trace coupling.

\subsection{Consistency conditions and GR limit}
\label{subsec:viability_background}

The General Relativity limit is recovered when $f(R,T_{\chi})\rightarrow R$ so that
$f_R\rightarrow 1$ and $f_T\rightarrow 0$, in which case
Eqs.~\eqref{eq:Friedmann_00}--\eqref{eq:Qchi_explicit}
reduce to the standard Friedmann system with a canonical scalar field.
For a well-defined background evolution, the theory must also avoid singular denominators,
in particular $\kappa^2\pm f_T\neq 0$ wherever Eqs.~\eqref{eq:chi_nonconservation} and
\eqref{eq:Qchi_explicit} are used, and it must satisfy the usual positivity requirement
$f_R>0$ to prevent an effective sign flip of the gravitational coupling \cite{DeFeliceTsujikawa2010fRReview,SotiriouFaraoni2010fRReview}.

\paragraph*{Viability and stability cuts.}
In addition to the GR limit and the absence of singular denominators, we impose
linear-viability conditions for a predictive late-time cosmology:
\begin{equation}
f_R>0,\qquad f_{RR}>0,\qquad K \equiv 1+\frac{f_T}{\kappa^2}>0,
\end{equation}
where $f_{RR}>0$ ensures Dolgov--Kawasaki stability in the curvature sector \cite{DolgovKawasaki2003,DeFeliceTsujikawa2010fRReview,SotiriouFaraoni2010fRReview} and
$K>0$ guarantees that the coefficient of $\delta\ddot{\chi}$ in the perturbed
Klein--Gordon equation (Eq.~46) remains positive, avoiding ghost-like kinetic
behavior of $\chi$ fluctuations.

\begin{figure}
    \centering
    \includegraphics[width=0.85\linewidth]{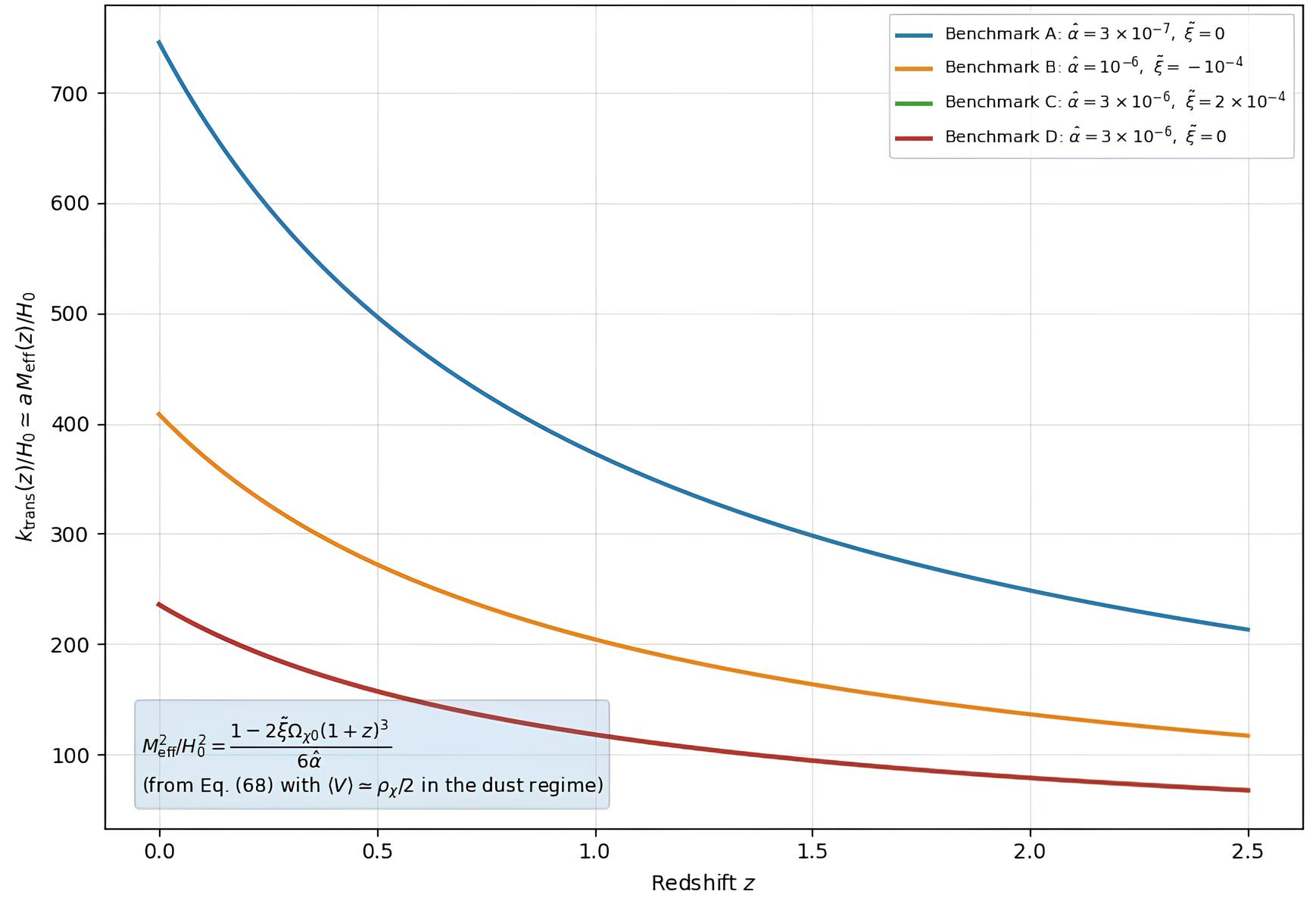}
    \caption{
Redshift evolution of the quasi-static transition scale
$k_{\rm trans}(z)\simeq a\,M_{\rm eff}(z)$ for viable benchmark points in the
$(\hat{\alpha},\tilde{\xi})$ plane, using the polynomial model
$f(R,T_\chi)=R+\alpha R^2+\xi R T_\chi+\beta T_\chi^2$ and the effective mass
expression in Eq.~\eqref{eq:Meff_Bcoef_model_safe}. The plotted curves use the dust-regime benchmark closure
$\langle V\rangle\simeq \rho_\chi/2$ (oscillating quadratic scalar), yielding
$M_{\rm eff}^2/H_0^2=[1-2\tilde{\xi}\Omega_{\chi0}(1+z)^3]/(6\hat{\alpha})$.
This figure identifies the scale at which scalaron-mediated modifications are
expected to transition in the quasi-static regime \cite{DeFeliceTsujikawa2010fRReview,Turner1983CoherentScalarOsc}.
} 
    \label{fig:2}
\end{figure}

\paragraph*{Effective dust regime of $\chi$.}
Although the selective trace source is defined using a canonical scalar field
$\chi$, the late-time large-scale observables considered here are most directly
interpreted when the dark sector clusters effectively as pressureless matter.
This is consistently realized, for instance, by an oscillating scalar in an
approximately quadratic potential in the regime $m_\chi\gg H$, for which
time-averaging yields $\langle w_\chi\rangle\simeq 0$ and hence
$\langle T_\chi\rangle \simeq -\langle\rho_\chi\rangle$ (up to corrections
suppressed by $H/m_\chi$) \cite{Turner1983CoherentScalarOsc,Marsh2016AxionCosmology,Hui2017UltralightScalarsDM}. In this limit, the standard growth equation used in
Sec.~\ref{sec:obsfe_inrence} applies, while the selective coupling effects enter through
$(f_T,\delta f_T)$ and the additional source terms derived in
Sec.~\ref{sec:perturbations}.

More precisely, the dust-like closure adopted here assumes that the background
field oscillates around a locally quadratic minimum,
$V(\chi)\simeq \tfrac12 m_\chi^2\chi^2$, with oscillation frequency $m_\chi$
much larger than the Hubble rate over the late-time redshift range relevant to
the linear observables considered. In this adiabatic/averaged regime, residual
pressure and gradient contributions are suppressed on sufficiently large scales \cite{Turner1983CoherentScalarOsc,Hui2017UltralightScalarsDM}.
Correspondingly, the perturbation-level mapping developed here is formulated for
the late-time large-scale linear regime in which the oscillation-averaged
effective dust approximation is valid. Within this regime, the standard
dust-based variable $f\sigma_8$ can be used consistently in the linear
observational mapping employed in Sec.~\ref{sec:obsfe_inrence}. The present
construction therefore targets the perturbation-level, late-time large-scale
sector, while preserving the selective-coupling modifications through
$(f_T,\delta f_T)$ and the extra source terms derived in
Sec.~\ref{sec:perturbations}.

\section{Linear cosmological perturbations}
\label{sec:perturbations}

\subsection{Newtonian gauge and perturbation variables}
We work with scalar perturbations around the flat FLRW background in Newtonian gauge \cite{MaBertschinger1995},
\begin{equation}
\mathrm{d}s^2 = -(1+2\Psi)\,\mathrm{d}t^2 + a^2(t)(1-2\Phi)\,\delta_{ij}\mathrm{d}x^i \mathrm{d}x^j,
\label{eq:newtonian_gauge_metric}
\end{equation}
where $\Psi(t,\mathbf{x})$ and $\Phi(t,\mathbf{x})$ are the Bardeen potentials.
The dark-matter field is decomposed as
\begin{equation}
\chi(t,\mathbf{x})=\bar{\chi}(t)+\delta\chi(t,\mathbf{x}),
\label{eq:chi_split}
\end{equation}
and similarly for all background quantities. We denote Fourier modes by
$\delta X(t,\mathbf{x})=\int \frac{\mathrm{d}^3k}{(2\pi)^3}\,\delta X(t,\mathbf{k})\,e^{i\mathbf{k}\cdot\mathbf{x}}$.

For later use, we define the total (visible + dark) matter density perturbation by
$\delta\rho \equiv \delta\rho_b+\delta\rho_{\rm rad}+\delta\rho_\chi$,
and introduce the velocity potentials $v_I$ (for each species $I$) through
$\delta u^{(I)}_i = \partial_i v_I$ at linear order.

\subsection{Perturbations of $T_{\chi}$ and of the coupling functions}
The trace $T_{\chi}$ defined in Eq.~\eqref{eq:Tchi_def} perturbs as
$T_{\chi}=\bar{T}_{\chi}+\delta T_{\chi}$.
For the canonical scalar field in Newtonian gauge, keeping only linear terms, one finds
\begin{equation}
\delta T_{\chi}
=2\dot{\bar{\chi}}\,\delta\dot{\chi}
-2\Psi\,\dot{\bar{\chi}}^{\,2}
-4V_{,\chi}\,\delta\chi ,
\label{eq:deltaTchi}
\end{equation}
where $V_{,\chi}\equiv \mathrm{d}V/\mathrm{d}\chi$ is evaluated on the background.

Since $f_R$ and $f_T$ depend on $(R,T_{\chi})$, their perturbations are
\begin{align}
\delta f_R &= f_{RR}\,\delta R + f_{RT}\,\delta T_{\chi}, \label{eq:deltafR}\\
\delta f_T &= f_{TR}\,\delta R + f_{TT}\,\delta T_{\chi}, \label{eq:deltafT}
\end{align}
where $f_{RR}\equiv \partial^2 f/\partial R^2$, $f_{RT}\equiv \partial^2 f/(\partial R\,\partial T_\chi)$,
and $f_{TT}\equiv \partial^2 f/\partial T_\chi^2$ are evaluated on the background.
The Ricci-scalar perturbation $\delta R$ is obtained from Eq.~\eqref{eq:newtonian_gauge_metric}
and will enter the linearized field equations through Eqs.~\eqref{eq:deltafR}--\eqref{eq:deltafT}.
For concreteness and parameter-level interpretability, we adopt the minimal polynomial realization
\begin{equation}
f(R,T_\chi)=R+\alpha R^2+\xi R T_\chi+\beta T_\chi^2 ,
\label{eq:model_f_poly}
\end{equation}
as an effective expansion around the GR limit. These relations also clarify the parameter roles
at linear order in this parametrization: $\alpha$ enters primarily through the curvature-response
sector (via $f_{RR}$ and the effective scalaron scale), $\xi$ mixes curvature and dark-sector trace
perturbations through $f_{RT}=f_{TR}$, and $\beta$ controls the leading trace-sector response
through $f_{TT}$. As a result, different observables probe different combinations of
$(\alpha,\xi,\beta)$: background information alone is generally insufficient to separate them,
whereas growth and lensing respond differently to the curvature and trace-coupling channels \cite{Koyama2016CosmoTestsMG,Ishak2019TestingGRCosmology,PogosianSilvestri2016SigmaMu}.

\subsection{Effective functions for growth and lensing}
A key objective is to connect the linearized system to perturbation-sensitive observables.
In Fourier space, and in the sub-horizon regime, the scalar sector can be parameterized by
effective modified-gravity functions $\mu(k,a)$ and $\eta(k,a)$ defined via \cite{PogosianSilvestri2016SigmaMu,Ishak2019TestingGRCosmology}
\begin{equation}
k^2\Psi \equiv -4\pi G\,a^2\,\mu(k,a)\,\rho_m\,\Delta_m,
\qquad
\eta(k,a)\equiv \frac{\Phi}{\Psi},
\label{eq:mu_eta_defs}
\end{equation}
where $\rho_m$ is the background total matter density (typically baryons plus the clustering
dark component), and $\Delta_m$ is the gauge-invariant comoving matter density contrast \cite{MaBertschinger1995}.

\paragraph*{No double counting between $\rho_m\Delta_m$ and $\Pi_{\rm DS}$.}
In our notation, $\rho_m\Delta_m$ denotes the standard comoving clustering
source entering the phenomenological definitions of $\mu$ and $\Sigma$,
including the clustering dark component in the effective dust regime.
The additional term $\Pi_{\rm DS}$ introduced in the QS Poisson constraint
(Eq.~60) collects only the selective trace-coupling contributions proportional
to $(f_T,\delta f_T)$ and vanishes in the GR limit $f_T\to 0$; therefore
$\Pi_{\rm DS}$ should be interpreted as an extra modified-gravity source,
not as a second counting of the matter density perturbation.

The Weyl (lensing) potential is controlled by $\Phi+\Psi$, and it is convenient to define \cite{PogosianSilvestri2016SigmaMu,Ishak2019TestingGRCosmology}
\begin{equation}
k^2(\Phi+\Psi)\equiv -8\pi G\,a^2\,\Sigma(k,a)\,\rho_m\,\Delta_m,
\qquad
\Sigma(k,a)\equiv \frac{\mu(k,a)}{2}\left[1+\eta(k,a)\right].
\label{eq:Sigma_def}
\end{equation}
In the following subsections we derive the full linearized equations from
Eq.~\eqref{eq:field_eqs_general} and identify $\mu(k,a)$ and $\eta(k,a)$ explicitly in terms
of the model functions and background evolution, enabling direct confrontation with
redshift-space distortions (via $f\sigma_8$) and lensing data.

\subsection{Perturbed dark-matter stress--energy}
\label{subsec:perturbed_dm}

For the canonical field $\chi$ defined in Eq.~\eqref{eq:Lchi}, the linear perturbations of
energy density, pressure, and momentum density in Newtonian gauge are \cite{MaBertschinger1995,Mukhanov2005PhysicalFoundations} 
\begin{align}
\delta\rho_{\chi} &= \dot{\bar{\chi}}\,\delta\dot{\chi} - \dot{\bar{\chi}}^{\,2}\Psi + V_{,\chi}\,\delta\chi,
\label{eq:delta_rho_chi}\\
\delta p_{\chi} &= \dot{\bar{\chi}}\,\delta\dot{\chi} - \dot{\bar{\chi}}^{\,2}\Psi - V_{,\chi}\,\delta\chi,
\label{eq:delta_p_chi}\\
\delta T^{(\chi)}_{0i} &= -\,\dot{\bar{\chi}}\,\partial_i\delta\chi
\quad\Rightarrow\quad
(\bar{\rho}_{\chi}+\bar{p}_{\chi})\,v_{\chi} = -\,\dot{\bar{\chi}}\,\delta\chi,
\label{eq:momentum_chi}
\end{align}
where $V_{,\chi}\equiv \mathrm{d}V/\mathrm{d}\chi$ is evaluated on the background.

\subsection{Linearized field equations and robust slip relation}
\label{subsec:linearized_core}

Define the geometric operator
\begin{equation}
\mathcal{E}_{\mu\nu}\equiv
f_R R_{\mu\nu}-\frac{1}{2}f\,g_{\mu\nu}
+\left(g_{\mu\nu}\Box-\nabla_{\mu}\nabla_{\nu}\right)f_R .
\label{eq:E_munu_def}
\end{equation}
The field equations \eqref{eq:field_eqs_general} can be written compactly as
\begin{equation}
\mathcal{E}_{\mu\nu}
=\kappa^2\left(T^{(b)}_{\mu\nu}+T^{(\mathrm{rad})}_{\mu\nu}+T^{(\chi)}_{\mu\nu}\right)
- f_T\,S_{\mu\nu},
\qquad
S_{\mu\nu}\equiv T^{(\chi)}_{\mu\nu}+\Theta^{(\chi)}_{\mu\nu}.
\label{eq:field_eq_compact}
\end{equation}
For canonical $\chi$, Eq.~\eqref{eq:TplusTheta_simple} implies the exact identity
\begin{equation}
S_{\mu\nu}=-\partial_{\mu}\chi\,\partial_{\nu}\chi .
\label{eq:Smunu_identity}
\end{equation}
At the background level one has $\bar{S}_{00}=-\dot{\bar{\chi}}^{\,2}$ and $\bar{S}_{0i}=\bar{S}_{ij}=0$.
At linear order this gives the particularly simple perturbations
\begin{equation}
\delta S_{00}=-2\dot{\bar{\chi}}\,\delta\dot{\chi},
\qquad
\delta S_{0i}=-\dot{\bar{\chi}}\,\partial_i\delta\chi,
\qquad
\delta S_{ij}=0 \ \ (\text{to first order}).
\label{eq:deltaS_components}
\end{equation}
Therefore, the linearized field equations take the schematic but exact form
\begin{equation}
\delta\mathcal{E}_{\mu\nu}
=\kappa^2\,\delta T^{\rm (tot)}_{\mu\nu}
-\bar{f}_T\,\delta S_{\mu\nu}
-\delta f_T\,\bar{S}_{\mu\nu},
\label{eq:lin_field_schematic}
\end{equation}
where $\delta T^{\rm (tot)}_{\mu\nu}$ includes baryons, radiation, and $\chi$.

A key robust consequence follows from the traceless spatial ($ij$) sector.
Since $\delta S_{ij}=0$ at linear order and the canonical scalar has no anisotropic stress,
the traceless $ij$ equation yields the gravitational slip sourced purely by $\delta f_R$ \cite{DeFeliceTsujikawa2010fRReview,SotiriouFaraoni2010fRReview},
\begin{equation}
\Phi-\Psi=\frac{\delta f_R}{\bar{f}_R}.
\label{eq:slip_relation_final}
\end{equation}
In the next subsection we will compute the explicit $00$ and $0i$ components of
Eq.~\eqref{eq:lin_field_schematic} in Fourier space and then extract $\mu(k,a)$ and $\eta(k,a)$
consistently with Eqs.~\eqref{eq:mu_eta_defs}--\eqref{eq:Sigma_def}.

\begin{figure}
    \centering
    \includegraphics[width=0.85\linewidth]{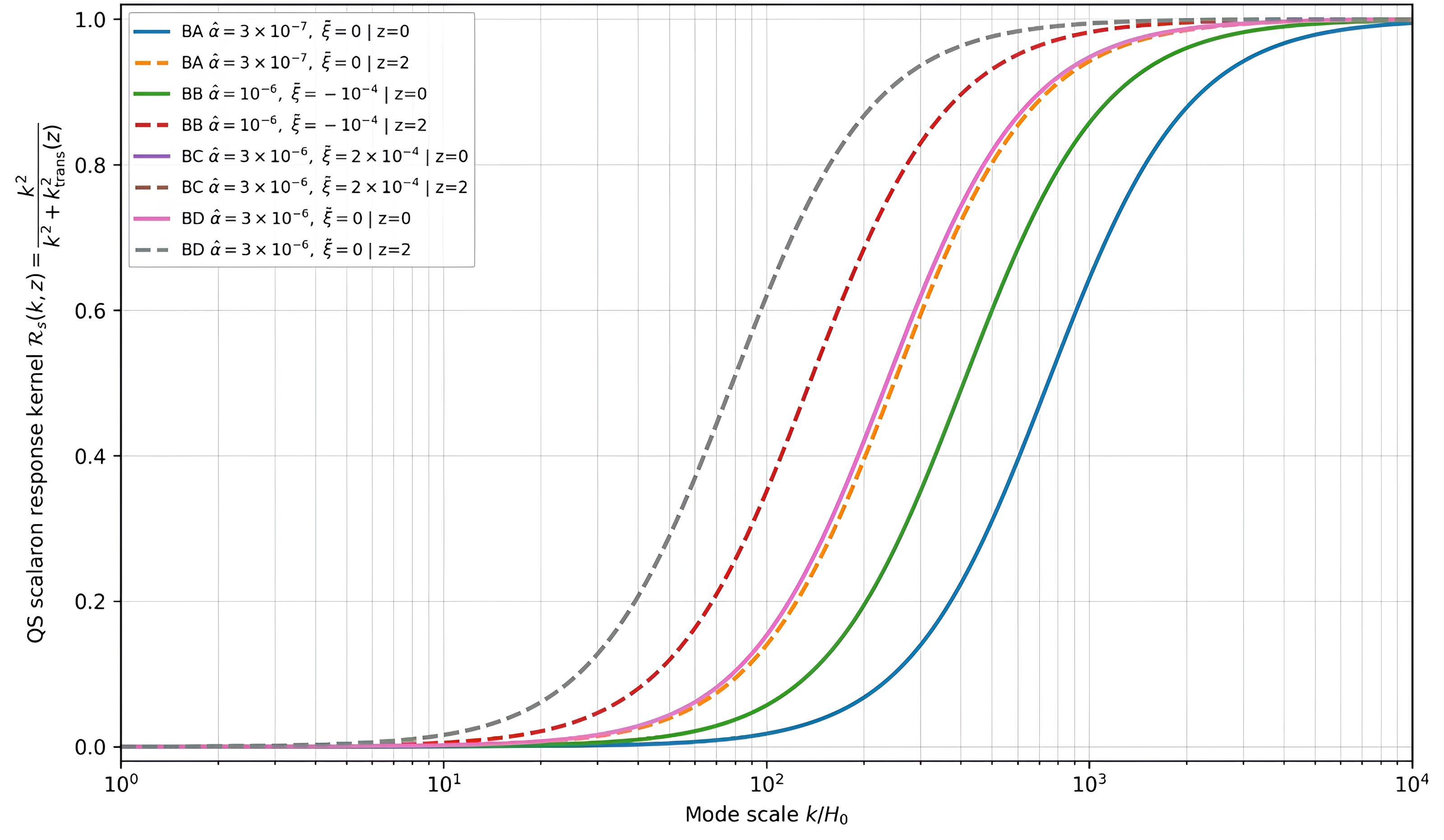}
    \caption{
Diagnostic quasi-static scalaron response kernel,
$\mathcal{R}_s(k,z)=k^2/[k^2+k_{\rm trans}^2(z)]$, for viable benchmark points
of the selective dark-sector $f(R,T_\chi)$ model, where
$k_{\rm trans}(z)\simeq a\,M_{\rm eff}(z)$ is determined by the effective mass
derived in Sec.~\ref{subsec:model_QS_trace}. Solid (dashed) curves denote $z=0$ ($z=2$). This plot isolates the scale at which scalaron-mediated modifications become active and
provides a theory-level bridge between the effective-mass sector and the
scale-dependent growth/lensing observables reconstructed in the quasi-static
regime.
}
    \label{fig:3}
\end{figure}

\subsection{Linearized modified Klein--Gordon equation}
\label{subsec:lin_KG}

The dark-matter field equation follows from varying the action with respect to $\chi$.
It is convenient to introduce the dimensionless coupling
\begin{equation}
B \equiv \frac{f_T}{\kappa^2},
\qquad
\mathcal{K}\equiv 1+B,
\qquad
\mathcal{C}\equiv 1+2B,
\label{eq:K_C_defs}
\end{equation}
so that the exact covariant equation of motion can be written as
\begin{equation}
\mathcal{K}\,\Box\chi
+ \nabla_{\mu}B\,\nabla^{\mu}\chi
-\mathcal{C}\,V_{,\chi}=0.
\label{eq:chi_covariant_eom}
\end{equation}
On the FLRW background this reproduces Eq.~\eqref{eq:chi_background_eom}.

Linearizing Eq.~\eqref{eq:chi_covariant_eom} around $\chi=\bar{\chi}+\delta\chi$ and
$B=\bar{B}+\delta B$ yields, to first order,
\begin{equation}
\bar{\mathcal{K}}\,\delta(\Box\chi)
+\delta\mathcal{K}\,\overline{\Box\chi}
+\nabla_{\mu}\delta B\,\nabla^{\mu}\bar{\chi}
+\nabla_{\mu}\bar{B}\,\nabla^{\mu}\delta\chi
+\delta g^{\mu\nu}\,(\partial_{\mu}\bar{B})(\partial_{\nu}\bar{\chi})
-\bar{\mathcal{C}}\,V_{,\chi\chi}\,\delta\chi
-\delta\mathcal{C}\,V_{,\chi}
=0,
\label{eq:lin_KG_compact}
\end{equation}
where $\delta\mathcal{K}=\delta B$ and $\delta\mathcal{C}=2\,\delta B$.

The coupling perturbation $\delta B$ is determined by the gravitational function through
\begin{equation}
\delta B=\frac{1}{\kappa^2}\,\delta f_T
=\frac{1}{\kappa^2}\left(f_{TR}\,\delta R + f_{TT}\,\delta T_{\chi}\right),
\label{eq:deltaB_from_f}
\end{equation}
with $\delta T_{\chi}$ given in Eq.~\eqref{eq:deltaTchi}.
On the FLRW background, $\delta g^{00}=2\Psi$ and
\begin{equation}
\delta g^{\mu\nu}(\partial_{\mu}\bar{B})(\partial_{\nu}\bar{\chi})
= 2\Psi\,\dot{\bar{B}}\,\dot{\bar{\chi}}.
\label{eq:delta_g_term_simplify}
\end{equation}

Finally, the perturbed d'Alembertian acting on $\chi$ is defined geometrically by
\begin{equation}
\delta(\Box\chi)
\equiv
\delta\!\left(g^{\mu\nu}\nabla_{\mu}\nabla_{\nu}\chi\right)
=
\bar{\Box}\,\delta\chi
+\delta g^{\mu\nu}\,\bar{\nabla}_{\mu}\bar{\nabla}_{\nu}\bar{\chi}
+\bar{g}^{\mu\nu}\,\delta\Gamma^{\rho}{}_{\mu\nu}\,\partial_{\rho}\bar{\chi},
\label{eq:deltaBox_def}
\end{equation}
which can be evaluated explicitly in Newtonian gauge using Eq.~\eqref{eq:newtonian_gauge_metric}.
In the next subsection we provide the explicit Fourier-space form and then close the system
for $(\Phi,\Psi,\delta\chi)$ to extract $\mu(k,a)$ and $\eta(k,a)$ consistently with
Eqs.~\eqref{eq:mu_eta_defs}--\eqref{eq:Sigma_def}.

\subsection{Explicit Newtonian-gauge form and Fourier-space equation for $\delta\chi$}
\label{subsec:lin_KG_explicit}

\subsubsection*{Explicit form of $\delta(\Box\chi)$}
Evaluating Eq.~\eqref{eq:deltaBox_def} for the Newtonian-gauge metric
\eqref{eq:newtonian_gauge_metric} and a homogeneous background $\bar{\chi}(t)$ gives
\begin{equation}
\delta(\Box\chi)
=
-\delta\ddot{\chi}
-3H\,\delta\dot{\chi}
-\frac{k^2}{a^2}\,\delta\chi
+\dot{\bar{\chi}}\left(\dot{\Psi}+3\dot{\Phi}\right)
+2\Psi\left(\ddot{\bar{\chi}}+3H\dot{\bar{\chi}}\right),
\label{eq:deltaBox_explicit}
\end{equation}
where $k$ is the comoving wavenumber in Fourier space.

\subsubsection*{Fourier-space linearized modified Klein--Gordon equation}
Using Eq.~\eqref{eq:deltaBox_explicit} in the compact linear equation
\eqref{eq:lin_KG_compact}, and keeping the definitions
$\mathcal{K}=1+B$, $\mathcal{C}=1+2B$ from Eq.~\eqref{eq:K_C_defs}, one obtains
\begin{align}
\bar{\mathcal{K}}\,\delta\ddot{\chi}
+\left(3H\bar{\mathcal{K}}+\dot{\bar{B}}\right)\delta\dot{\chi}
+\left(\bar{\mathcal{K}}\frac{k^2}{a^2}+\bar{\mathcal{C}}\,V_{,\chi\chi}\right)\delta\chi
&=
\bar{\mathcal{K}}\,\dot{\bar{\chi}}\left(\dot{\Psi}+3\dot{\Phi}\right)
-2\bar{\mathcal{C}}\,V_{,\chi}\,\Psi
-\dot{\bar{\chi}}\,\delta\dot{B}
+\delta B\,\overline{\Box\chi}
-2V_{,\chi}\,\delta B,
\label{eq:lin_KG_Fourier}
\end{align}
where
\begin{equation}
\overline{\Box\chi}=-(\ddot{\bar{\chi}}+3H\dot{\bar{\chi}}),
\label{eq:Boxchi_background}
\end{equation}
and the coupling perturbation $\delta B$ is
\begin{equation}
\delta B=\frac{1}{\kappa^2}\left(f_{TR}\,\delta R + f_{TT}\,\delta T_{\chi}\right),
\qquad
\delta T_{\chi}
=2\dot{\bar{\chi}}\,\delta\dot{\chi}-2\Psi\,\dot{\bar{\chi}}^{\,2}-4V_{,\chi}\,\delta\chi,
\label{eq:deltaB_deltaTchi_repeat}
\end{equation}
with all background-dependent coefficients evaluated on $(\bar{R},\bar{T}_{\chi})$.
Equation \eqref{eq:lin_KG_Fourier} closes the $\delta\chi$ dynamics once $\delta R$ is specified
from the metric perturbations.

\subsection{Eliminating $\delta R$ and closing the system}
\label{subsec:deltaR_eliminate}

A practical advantage of the $(R,T_{\chi})$ dependence is that $\delta R$ never needs to be
written explicitly in terms of $(\Phi,\Psi)$ in order to close the system.
From Eqs.~\eqref{eq:deltafR}--\eqref{eq:deltafT} we have, at linear order,
\begin{equation}
\delta f_R = f_{RR}\,\delta R + f_{RT}\,\delta T_{\chi},
\qquad
\delta f_T = f_{TR}\,\delta R + f_{TT}\,\delta T_{\chi},
\label{eq:deltafR_deltafT_pair}
\end{equation}
with $f_{RT}=f_{TR}$.
We assume $f_{RR}\neq 0$ so that $\delta R$ can be eliminated in favor of $\delta f_R$; the
special case $f_{RR}=0$ can be treated separately.
Solving the first relation for $\delta R$ gives
\begin{equation}
\delta R = \frac{1}{f_{RR}}\left(\delta f_R - f_{RT}\,\delta T_{\chi}\right).
\label{eq:deltaR_from_deltafR}
\end{equation}
Substituting Eq.~\eqref{eq:deltaR_from_deltafR} into $\delta f_T$ yields an expression that depends
only on $\delta f_R$ and $\delta T_{\chi}$,
\begin{equation}
\delta f_T
=
\frac{f_{RT}}{f_{RR}}\,\delta f_R
+\left(f_{TT}-\frac{f_{RT}^2}{f_{RR}}\right)\delta T_{\chi}.
\label{eq:deltafT_closed}
\end{equation}
Therefore the coupling perturbation $\delta B=\delta f_T/\kappa^2$ becomes
\begin{equation}
\delta B
=
\frac{1}{\kappa^2}\frac{f_{RT}}{f_{RR}}\,\delta f_R
+\frac{1}{\kappa^2}\left(f_{TT}-\frac{f_{RT}^2}{f_{RR}}\right)\delta T_{\chi}.
\label{eq:deltaB_closed}
\end{equation}

Using the robust slip relation \eqref{eq:slip_relation_final},
\begin{equation}
\delta f_R = \bar{f}_R(\Phi-\Psi),
\label{eq:deltafR_from_slip}
\end{equation}
Eq.~\eqref{eq:deltaB_closed} can be expressed entirely in terms of $(\Phi,\Psi)$ and $\delta T_{\chi}$.
For reference, the canonical-field trace perturbation is
\begin{equation}
\delta T_{\chi}
=2\dot{\bar{\chi}}\,\delta\dot{\chi}
-2\Psi\,\dot{\bar{\chi}}^{\,2}
-4V_{,\chi}\,\delta\chi .
\label{eq:deltaTchi_repeat2}
\end{equation}

\subsection{Trace equation for the scalaron perturbation $\delta f_R$}
\label{subsec:trace_scalar}

Taking the trace of the field equations \eqref{eq:field_eqs_general} gives the exact scalar equation
\begin{equation}
3\Box f_R + f_R R - 2f
=
\kappa^2 T^{\rm (tot)} - f_T\,S,
\qquad
S\equiv g^{\mu\nu}(T^{(\chi)}_{\mu\nu}+\Theta^{(\chi)}_{\mu\nu})=-(\partial\chi)^2,
\label{eq:trace_exact}
\end{equation}
where $T^{\rm (tot)}$ is the trace of the total stress-energy tensor (baryons + radiation + $\chi$).
Linearizing Eq.~\eqref{eq:trace_exact} yields
\begin{equation}
3\,\delta(\Box f_R)
+\bar{R}\,\delta f_R
-\bar{f}_R\,\delta R
-2\bar{f}_T\,\delta T_{\chi}
=
\kappa^2\,\delta T^{\rm (tot)}
-\bar{f}_T\,\delta S
-\delta f_T\,\bar{S},
\label{eq:trace_linear}
\end{equation}
with $\delta R$ eliminated by Eq.~\eqref{eq:deltaR_from_deltafR} and $\delta f_T$ given by
Eq.~\eqref{eq:deltafT_closed}. For the canonical field,
\begin{equation}
\bar{S}=\dot{\bar{\chi}}^{\,2},
\qquad
\delta S = 2\dot{\bar{\chi}}\,\delta\dot{\chi}-2\Psi\,\dot{\bar{\chi}}^{\,2}.
\label{eq:S_background_pert}
\end{equation}

Finally, the perturbed d'Alembertian acting on $f_R$ is defined analogously to
Eq.~\eqref{eq:deltaBox_def} as
\begin{equation}
\delta(\Box f_R)
=
\bar{\Box}\,\delta f_R
+\delta g^{\mu\nu}\,\bar{\nabla}_{\mu}\bar{\nabla}_{\nu}\bar{f}_R
+\bar{g}^{\mu\nu}\,\delta\Gamma^{\rho}{}_{\mu\nu}\,\partial_{\rho}\bar{f}_R,
\label{eq:deltaBox_fR_def}
\end{equation}
where $\bar{f}_R=\bar{f}_R(t)$ in the background.

Equations \eqref{eq:lin_KG_Fourier}, \eqref{eq:trace_linear}, and \eqref{eq:slip_relation_final},
together with Eq.~\eqref{eq:deltaB_closed}, provide a gauge-consistent closure once supplemented by
the $00$ and $0i$ constraint equations for the metric potentials.

\subsection{Quasi-static sub-horizon limit and observable reconstruction}
\label{subsec:QS_mu_eta_safe}

We now specialize to the quasi-static (QS) sub-horizon regime \cite{DeFeliceTsujikawa2010fRReview,Koyama2016CosmoTestsMG},
\begin{equation}
\frac{k^2}{a^2}\gg H^2,
\qquad
\left|\dot{\Phi}\right|,\left|\dot{\Psi}\right|\ll \frac{k}{a}\left|\Phi\right|,\frac{k}{a}\left|\Psi\right|,
\qquad
\left|\delta\dot f_R\right|\ll \frac{k}{a}\left|\delta f_R\right|,
\label{eq:QS_assumptions_safe}
\end{equation}
retaining the leading spatial-gradient terms that control growth and lensing.

\subsubsection*{QS constraints for the metric potentials}
In the QS limit, the linearized field equations \eqref{eq:lin_field_schematic} reduce to a
Poisson-type constraint for $\Phi$ with two distinct sources:
(i) the standard clustering source $\rho_m\Delta_m$ and
(ii) a dark-sector selective contribution induced by $(f_T,\delta f_T)$ and $\delta S_{00}$.
We write the leading-gradient constraint in the compact form
\begin{equation}
\frac{k^2}{a^2}\Phi \simeq
-\,\frac{\kappa^2}{2\bar f_R}\,\rho_m\Delta_m
-\,\frac{1}{2\bar f_R}\,\Pi_{\rm DS}(k,a)
-\,\frac{k^2}{2a^2}\frac{\delta f_R}{\bar f_R},
\label{eq:QS_Phi_compact}
\end{equation}
where $\Pi_{\rm DS}$ collects the dark-sector selective terms that vanish in the GR limit
and is given by
\begin{equation}
\Pi_{\rm DS}(k,a)\equiv
\dot{\bar\chi}^{\,2}\,\delta f_T-\bar f_T\,\delta S_{00},
\qquad
\delta S_{00}=-2\dot{\bar\chi}\,\delta\dot\chi.
\label{eq:PiDS_def}
\end{equation}
The coupling perturbation $\delta f_T$ is eliminated by Eq.~\eqref{eq:deltafT_closed},
so that $\Pi_{\rm DS}$ is expressed entirely in terms of $\delta f_R$ and $(\delta\chi,\delta\dot\chi)$.

The gravitational slip remains exact at linear order,
\begin{equation}
\Phi-\Psi=\frac{\delta f_R}{\bar f_R},
\label{eq:slip_QS_safe}
\end{equation}
so that $\Psi$ is obtained from $\Phi$ once $\delta f_R$ is known.

\subsubsection*{QS trace equation for $\delta f_R$}
In the same QS limit, the trace equation \eqref{eq:trace_linear} becomes an algebraic relation
that determines $\delta f_R$ up to the source terms. Keeping the dominant gradient term
$\delta(\Box f_R)\simeq -(k^2/a^2)\delta f_R$ and eliminating $\delta R$ and $\delta f_T$ using
Eqs.~\eqref{eq:deltaR_from_deltafR} and \eqref{eq:deltafT_closed}, we obtain the schematic QS form
\begin{equation}
\left(\frac{k^2}{a^2}+m_f^2(a)\right)\delta f_R
\simeq
\mathcal{S}_{\rm tot}(k,a)+\mathcal{S}_{\rm DS}(k,a),
\label{eq:QS_trace_schematic}
\end{equation}
where $m_f^2(a)$ is the effective mass term for the scalaron perturbation $\delta f_R$ and
$\mathcal{S}_{\rm tot}$ and $\mathcal{S}_{\rm DS}$ denote, respectively, the standard trace source
(from $\delta T^{\rm (tot)}$) and the additional dark-sector selective sources involving
$\delta T_{\chi}$ and $\delta S$.
Their explicit expressions follow uniquely once a concrete model $f(R,T_{\chi})$ is specified,
because they depend only on the background derivatives
$(\bar f_R,\bar f_{RR},\bar f_{RT},\bar f_T,\bar f_{TT})$ and on the perturbations
$(\delta\chi,\delta\dot\chi,\Phi,\Psi)$.

\subsubsection*{Reconstruction of $\mu(k,a)$, $\eta(k,a)$ and $\Sigma(k,a)$}
Given $\delta f_R$ from Eq.~\eqref{eq:QS_trace_schematic}, the potentials are obtained from
Eqs.~\eqref{eq:QS_Phi_compact} and \eqref{eq:slip_QS_safe}, and the observable functions are then
read off from their defining relations
\eqref{eq:mu_eta_defs}--\eqref{eq:Sigma_def}:
\begin{equation}
\mu(k,a)= -\frac{k^2\Psi}{4\pi G a^2\,\rho_m\Delta_m},
\qquad
\eta(k,a)=\frac{\Phi}{\Psi},
\qquad
\Sigma(k,a)= -\frac{k^2(\Phi+\Psi)}{8\pi G a^2\,\rho_m\Delta_m}.
\label{eq:mu_eta_Sigma_reconstruct}
\end{equation}
In the GR limit $\bar f_R\to 1$ and $\Pi_{\rm DS}\to 0$ and $\delta f_R\to 0$, one recovers
$\mu=\eta=\Sigma=1$.

\subsection{QS scalaron equation and effective mass}
\label{subsec:model_QS_trace}

In the QS sub-horizon regime, the trace equation provides an algebraic relation for the
scalaron perturbation $\delta f_R$. Keeping the dominant gradient contribution
$\delta(\Box f_R)\simeq -(k^2/a^2)\delta f_R$ and eliminating $\delta R$ and $\delta f_T$ using
Eqs.~\eqref{eq:deltaR_from_deltafR} and \eqref{eq:deltafT_closed}, one may cast the result in the
compact form
\begin{equation}
\left(\frac{k^2}{a^2}+M_{\rm eff}^2(a)\right)\delta f_R
\simeq
-\frac{1}{3}\left[
\kappa^2\,\delta T^{\rm (tot)}
-\bar f_T\,\delta S
-\mathcal{B}(a)\,\delta T_{\chi}
\right],
\label{eq:QS_deltafR_model_safe}
\end{equation}
where the effective mass and the explicit $\delta T_{\chi}$-source coefficient are determined only by
background derivatives of $f(R,T_\chi)$:
\begin{equation}
M_{\rm eff}^2(a)\equiv
\frac{\bar f_R-\bar R\,\bar f_{RR}-\bar S\,\bar f_{RT}}{3\bar f_{RR}}
=
\frac{1}{3}\left[
\frac{\bar f_R-\bar S\,\bar f_{RT}}{\bar f_{RR}}-\bar R
\right],
\qquad
\mathcal{B}(a)\equiv
\bar f_R\,\frac{\bar f_{RT}}{\bar f_{RR}}
-2\bar f_T
+\bar S\left(\bar f_{TT}-\frac{\bar f_{RT}^2}{\bar f_{RR}}\right).
\label{eq:Meff_Bcoef_model_safe}
\end{equation}
For the canonical field, \(\bar S=\dot{\bar\chi}^{\,2}\) and \(\bar T_\chi=\bar S-4V(\bar\chi)\),
hence the coefficient \(\mathcal{B}(a)\) and the explicit \(\delta T_\chi\)-source contribution in
Eq.~\eqref{eq:QS_deltafR_model_safe} are fully determined by background derivatives of \(f(R,T_\chi)\)
together with the canonical scalar-sector perturbations under the QS sub-horizon approximation.

\begin{figure}[!t]
  \centering
  \includegraphics[width=\linewidth]{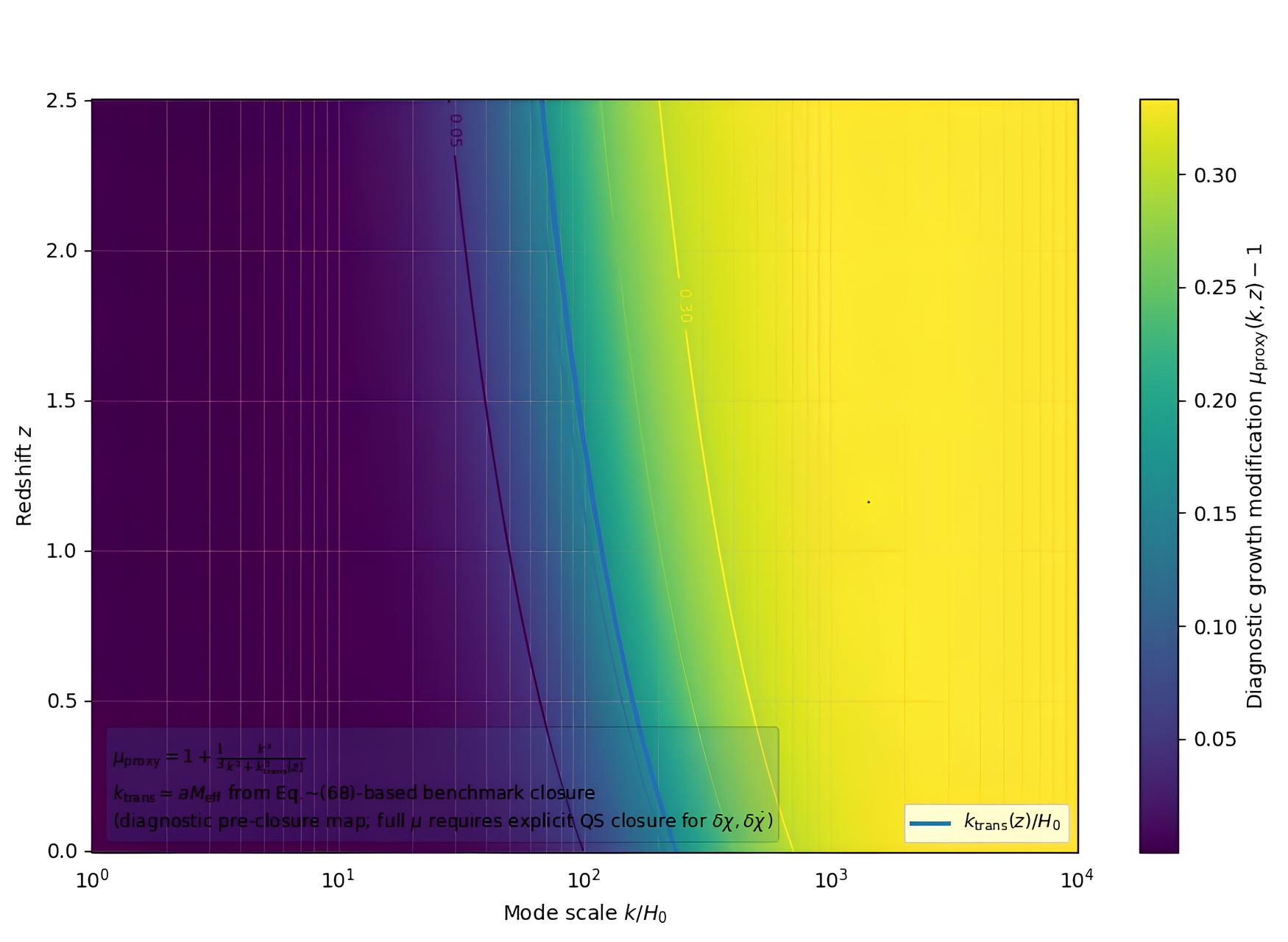} 
  \caption{\textbf{Diagnostic QS growth-modification map and transition scale.}
  Two-dimensional map of the diagnostic growth modification
  \(\mu_{\mathrm{proxy}}(k,z)-1\) in the quasi-static sub-horizon regime,
  displayed over mode scale \(k/H_0\) and redshift \(z\).
  The proxy \(\mu_{\mathrm{proxy}}(k,z)\) is constructed from the QS algebraic closure for the scalaron
  perturbation in Eq.~\eqref{eq:QS_deltafR_model_safe} with the background-defined coefficients
  \(M_{\rm eff}(a)\) and \(\mathcal{B}(a)\) given in Eq.~\eqref{eq:Meff_Bcoef_model_safe}.
  The solid curve shows the evolving transition locus \(k_{\rm trans}(z)/H_0\), illustrating that
  the onset of scale-dependent growth departures tracks the effective-mass scale
  \(k_{\rm trans}(z)\simeq aM_{\rm eff}(z)\).
  This figure therefore provides a compact pre-closure diagnostic of where, in the \((k,z)\) plane,
  the model predicts the strongest QS growth response.}
  \label{fig:4}
\end{figure}
\FloatBarrier

\begin{equation}
\delta S = 2\dot{\bar\chi}\,\delta\dot\chi-2\Psi\,\dot{\bar\chi}^{\,2},
\qquad
\delta T_{\chi}=\delta S-4V_{,\chi}\,\delta\chi
=
2\dot{\bar{\chi}}\,\delta\dot{\chi}
-2\Psi\,\dot{\bar{\chi}}^{\,2}
-4V_{,\chi}\,\delta\chi .
\label{eq:deltaS_deltaTchi_model_safe}
\end{equation}

For the specific polynomial choice in Eq.~\eqref{eq:model_f_poly}, using
$\bar f_{RR}=2\alpha$, $\bar f_{RT}=\xi$, $\bar f_{TT}=2\beta$ and
$\bar f_R=1+2\alpha\bar R+\xi\bar T_\chi$, one obtains the explicit background functions
\begin{equation}
M_{\rm eff}^2(a)=\frac{1}{3}\left[\frac{1+2\alpha\bar R+\xi\bar T_\chi-\xi\bar S}{2\alpha}-\bar R\right]
=\frac{1}{3}\left[\frac{1+\xi(\bar T_\chi-\bar S)}{2\alpha}\right]
=\frac{1-4\xi V(\bar\chi)}{6\alpha},
\label{eq:Meff_poly_explicit}
\end{equation}
and
\begin{equation}
\mathcal{B}(a)=
\left(1+2\alpha\bar R+\xi\bar T_\chi\right)\frac{\xi}{2\alpha}
-2\left(\xi\bar R+2\beta\bar T_\chi\right)
+\dot{\bar\chi}^{\,2}\left(2\beta-\frac{\xi^2}{2\alpha}\right).
\label{eq:Bcoef_poly_explicit}
\end{equation}
Equation \eqref{eq:QS_deltafR_model_safe}, together with the QS constraints
\eqref{eq:QS_Phi_compact} and the slip relation \eqref{eq:slip_QS_safe}, enables a fully explicit
reconstruction of $\mu(k,a)$, $\eta(k,a)$, and $\Sigma(k,a)$ once the perturbations
$(\delta\chi,\delta\dot\chi)$ are related to $\rho_m\Delta_m$ via the QS limit of
Eq.~\eqref{eq:lin_KG_Fourier} with $\delta B$ from Eq.~\eqref{eq:deltaB_from_f}.

\section{Observational signatures and inference strategy}
\label{sec:obsfe_inrence}

\subsection{Background expansion observables}
\label{subsec:background_obs}

Although we work in natural units ($c=\hbar=1$) in the theoretical sections, in this section we
restore $c$ when expressing observational distances in conventional units.
Assuming spatial flatness, we compute
\begin{equation}
D_H(z)=\frac{c}{H(z)},\qquad
D_M(z)=c\int_0^z \frac{dz'}{H(z')},\qquad
D_L(z)=(1+z)\,D_M(z),
\label{eq:distances_def}
\end{equation}
and the corresponding distance modulus for Type Ia supernovae \cite{Hogg1999DistanceMeasures},
\begin{equation}
\mu_{\rm SN}(z)=5\log_{10}\!\left[\frac{D_L(z)}{{\rm Mpc}}\right]+25.
\label{eq:mu_SN_def}
\end{equation}
For BAO analyses we adopt the standard volume-averaged distance
\begin{equation}
D_V(z)=\left[z\,D_M^2(z)\,D_H(z)\right]^{1/3},
\label{eq:DV_def}
\end{equation}
and/or the anisotropic pair $(D_M(z)/r_d,\;D_H(z)/r_d)$, where $r_d$ is the sound horizon at
the drag epoch computed consistently within the same background cosmology \cite{Eisenstein2005BAO,Percival2007BAO}.

\subsection{Linear growth and lensing in the QS regime}
\label{subsec:growth_lensing}

On sub-horizon scales where the QS approximation holds, clustering and lensing are controlled by the
effective functions $\mu(k,a)$ and $\Sigma(k,a)$ defined in
Eqs.~\eqref{eq:mu_eta_defs} and \eqref{eq:Sigma_def} \cite{DeFeliceTsujikawa2010fRReview,Koyama2016CosmoTestsMG,Ishak2019TestingGRCosmology}. For each Fourier mode $k$, the growth of the
comoving density contrast $\Delta_m(k,a)$ can be written as
\begin{equation}
\Delta_m'' + \left(2+\frac{H'}{H}\right)\Delta_m'
-\frac{3}{2}\,\Omega_m(a)\,\mu(k,a)\,\Delta_m=0,
\label{eq:growth_QS}
\end{equation}
where primes denote derivatives with respect to $\ln a$ and
$\Omega_m(a)\equiv \kappa^2\rho_m/(3H^2)$.
Redshift-space distortions constrain $f\sigma_8(z)$ with
\begin{equation}
f(z)\equiv \frac{d\ln \Delta_m}{d\ln a},\qquad
f\sigma_8(z)=f(z)\,\sigma_8(z),
\label{eq:fs8_def}
\end{equation}
while weak lensing and CMB lensing constrain the Weyl potential through $\Sigma(k,a)$,
\begin{equation}
k^2(\Phi+\Psi) = -8\pi G a^2\,\Sigma(k,a)\,\rho_m\Delta_m.
\label{eq:Weyl_Sigma}
\end{equation}
In our framework, $\mu$, $\eta$ and $\Sigma$ are reconstructed from
Eqs.~\eqref{eq:QS_Phi_compact}, \eqref{eq:slip_QS_safe}, and \eqref{eq:mu_eta_Sigma_reconstruct},
using $\delta f_R$ obtained from the QS trace equation \eqref{eq:QS_deltafR_model_safe}
and the model derivatives in Sec.~\ref{sec:framework}.

\begin{figure}[H]
  \centering
  \includegraphics[width=\linewidth]{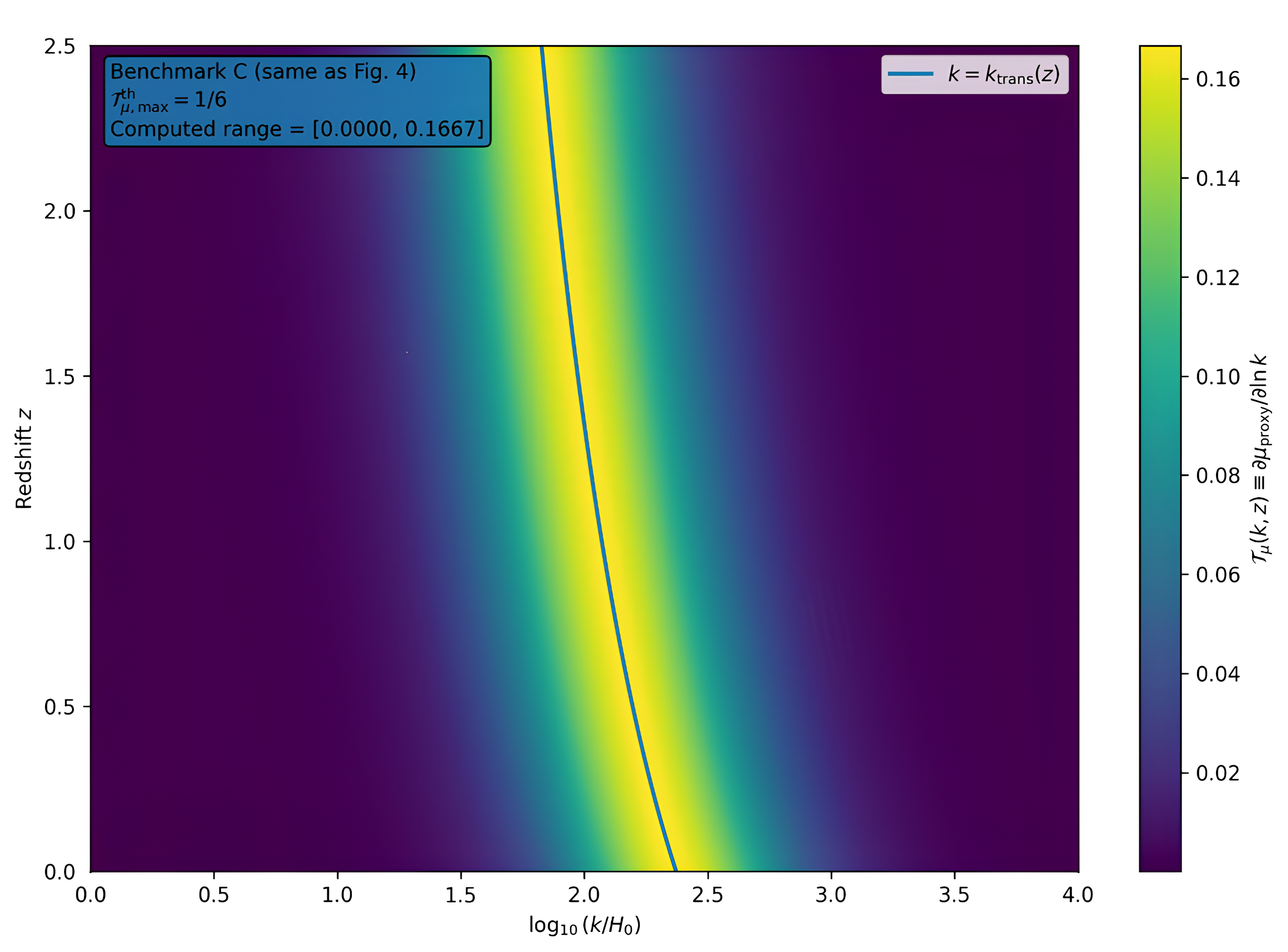}
  \caption{\textbf{Transition-sensitivity ridge of the quasi-static (QS) growth diagnostic.}
  Heat map of the logarithmic response measure
  $\mathcal{T}_{\mu}(k,z)\equiv \partial \mu_{\mathrm{proxy}}(k,z)/\partial\ln k$
  across the \((k,z)\) plane (horizontal axis: $\log_{10}(k/H_0)$; vertical axis: redshift \(z\)),
  for the same Benchmark~C setup adopted in the preceding analysis.
  The bright ridge delineates the scale band where the reconstructed QS growth response becomes
  maximally sensitive to mode scale, marking the effective transition between GR-like behavior and
  the modified-gravity scalar sector in the QS limit.
  The solid curve overlays the predicted transition locus \(k=k_{\mathrm{trans}}(z)\),
  highlighting that the strongest scale sensitivity tracks the evolving QS transition scale inferred
  from the effective-mass sector.
  The color bar reports the computed range of \(\mathcal{T}_{\mu}\) for this benchmark.}
  \label{fig:5}
\end{figure}
\FloatBarrier

\noindent Figure~\ref{fig:5} visualizes where the QS reconstruction becomes most
scale-sensitive in the \((k,z)\) plane, and shows that the resulting sensitivity ridge follows the
transition locus \(k_{\mathrm{trans}}(z)\) identified from the effective-mass sector.

\subsection{Data vector design for multi-probe inference}
\label{subsec:data_vector}

We define an implementation-ready multi-probe data-vector structure designed to separately test
(i) the background expansion and (ii) the growth and lensing sector:
\begin{equation}
\mathcal{D}=\left\{
{\rm BAO},\ {\rm SN},\ f\sigma_8,\ {\rm WL},\ {\rm CMB\ lensing}
\right\}.
\label{eq:data_vector}
\end{equation}
Geometric probes (BAO+SN) mainly constrain $H(z)$ and distances, while late-time clustering and lensing
probe the scale and time dependence of $\mu(k,a)$ and $\Sigma(k,a)$ in the linear regime.
As representative contemporary choices (included here only as concrete examples of the probe classes
entering the inference design), one may consider DESI DR2 BAO measurements, the Pantheon+ SN compilation,
DES Y6 weak-lensing and clustering products, and ACT DR6 CMB lensing products.

\subsection{Effective-likelihood construction and sampling blueprint}
\label{subsec:likelihood_sampling}

Assuming independent Gaussian likelihoods (or using published covariances when available), the
total log-likelihood can be written in the standard form
\begin{equation}
-2\ln\mathcal{L}_{\rm tot}
=
\sum_X
\left(\mathbf{d}_X-\mathbf{t}_X(\boldsymbol{\theta})\right)^{\!\rm T}
\mathbf{C}_X^{-1}
\left(\mathbf{d}_X-\mathbf{t}_X(\boldsymbol{\theta})\right),
\label{eq:likelihood_total}
\end{equation}
where $\mathbf{d}_X$ and $\mathbf{C}_X$ denote the data vector and covariance for probe $X$,
and $\mathbf{t}_X(\boldsymbol{\theta})$ are the corresponding theory predictions.
This expression defines the effective-likelihood layer used to map the present perturbation-level formulation
to an implementation-ready inference pipeline.
Within the scope of this work, it is used as a sampling blueprint and parameter-structure map for translated
linear-regime consistency diagnostics, rather than as a completed Boltzmann-level survey fit.

\subsection{Stability cuts and theory priors}
\label{subsec:stability_priors}

To ensure viability throughout the late-time redshift range targeted by the observational mapping considered here, we impose hard stability cuts:
\begin{align}
&\bar f_R(a)>0, \label{eq:prior_fR_pos}\\
&\alpha>0, \label{eq:prior_alpha_pos}\\
&|\bar f_T(a)|\ll \kappa^2 \ \ \text{at late times}, \label{eq:prior_fT_small}\\
&M_{\rm eff}^2(a)>0, \label{eq:prior_Meff_pos}
\end{align}
together with the non-singularity conditions summarized in Sec.~\ref{subsec:viability_background}.
The priors $\Pi(\boldsymbol{\theta})$ are taken broad and minimally informative on cosmological parameters,
while $(\alpha,\xi,\beta)$ are sampled in log-space over ranges that keep the QS reconstruction valid
and avoid strong-coupling regimes.

\paragraph{Scope of the present stability analysis.}
The stability conditions imposed above are intended as \emph{linear-viability} requirements for the late-time regime considered in this work, i.e.\ for background evolution, quasi-static
perturbations, and the effective-likelihood mapping to growth/lensing observables. They do not
constitute a full non-linear stability proof of the theory. A dedicated non-linear treatment
(including mode coupling and small-scale structure formation effects) is beyond the scope of the present analysis.

\section{Observational inputs and likelihood}
\label{sec:data_like}

In the absence of a dedicated Boltzmann-code implementation for the present
$f(R,T_\chi)$ realization (including the selective dark-sector coupling),
we adopt a conservative \emph{constraint-translation} strategy:
we anchor the late-time, linear-scale departures from GR to current bounds on
phenomenological modified-gravity functions and on the present-day scalar
response in viable $f(R)$-like scenarios.

By construction, the present inference layer is restricted to late-time linear
(or quasi-static linear) observables. We do not model non-linear structure formation,
halo-scale clustering, or screening/environmental effects in this work, since these
require a dedicated Boltzmann implementation coupled to a calibrated non-linear pipeline.
Accordingly, all translated constraints reported below should be interpreted as
conservative linear-regime benchmarks rather than full non-linear parameter bounds.
Theoretical viability is enforced separately through the stability cuts and theory priors
introduced in Sec.~\ref{subsec:stability_priors}.

This choice should be viewed as a conservative intermediate step between a formal
theoretical formulation and a full Boltzmann-level data analysis: while it does not provide
direct spectrum-level constraints, it yields a non-trivial and falsifiable late-time benchmark
through translated bounds on $(f_{R0},\mu_0,\Sigma_0)$ combined with theory-viability cuts \cite{YildizKaykiGudekli2026ISCausalProxy}.

\subsection{Phenomenological bounds on $\mu$ and $\Sigma$}
\label{subsec:muSigma_bounds}

We use updated constraints on the $(\mu,\Sigma)$ parameterization inferred from joint CMB and late-time probes.
In particular, Andrade et al.\ report (for a combined analysis including ACT, WMAP and late-time datasets)
\begin{equation}
\mu_0-1 = 0.02 \pm 0.19,\qquad
\Sigma_0-1 = 0.021 \pm 0.068,
\label{eq:muSigma_bounds}
\end{equation}
at 68\% confidence level \cite{Andrade2024MuSigma}. These constraints are consistent with GR and provide a robust
benchmark for allowing (or excluding) percent-level modifications to clustering and lensing at $z\simeq 0$.

\subsection{Benchmark bound on the present-day scalar response}
\label{subsec:fR0_bound}

For viable $f(R)$-type theories, an informative late-time handle is the present-day quantity
$f_{R0}\equiv f_R(R_0)-1$, which controls the effective gravitational strength on linear scales (in the quasi-static regime, $\mu\simeq 1/f_R$ up to scale-dependent corrections; see, e.g., \cite{DeFeliceTsujikawa2010fRReview}).
As a representative benchmark from galaxy clustering $\times$ CMB lensing cross-correlations, Kou et al.\ report
95\% limits on $|f_{R0}|$ at the level
\begin{equation}
\log_{10}\!\left|f_{R0}\right| < -4.18 \quad (95\%~\mathrm{CL}),
\label{eq:fR0_bound}
\end{equation}
with the precise number depending mildly on analysis choices \cite{Kou2023fR0}.
We adopt Eq.~\eqref{eq:fR0_bound} as a conservative cap on the present-day scalar response for any model whose
late-time linear phenomenology reduces to an $f_R$-controlled effective Newton constant.

\subsection{Effective likelihood layer}
\label{subsec:effective_like}

\paragraph{Parameter response map and degeneracy considerations.}
At the level of linear perturbations, the three coefficients play qualitatively different roles.
The curvature coefficient $\alpha$ controls the scalaron response scale (through the effective
curvature sector), the mixing coefficient $\xi$ controls the strength of the selective
curvature--dark-sector trace coupling, and $\beta$ governs the leading trace-sector self-interaction.
Consequently, late-time constraints often act primarily on parameter \emph{combinations} rather than
on each coefficient separately (e.g.\ through the present-day scalar response). In particular, the
translated $f_{R0}$ bound constrains a combination of $(\alpha,\xi)$ at $z=0$, while the
$(\mu_0,\Sigma_0)$ consistency layer provides an additional phenomenological handle on the relative
impact of curvature and lensing-sector modifications. A full multi-probe degeneracy-breaking analysis
using growth and lensing observables across redshift is deferred to future work with a dedicated
Boltzmann-code implementation; here we retain a conservative effective-likelihood layer designed to
preserve parameter interpretability while avoiding over-claiming.

We implement Eq.~\eqref{eq:fR0_bound} as a Gaussian pseudo-likelihood for $f_{R0}$ centered on GR,
\begin{equation}
\mathcal{L}_{f_{R0}} \propto \exp\!\left[-\frac{(f_{R0}-0)^2}{2\sigma_{f_{R0}}^2}\right],
\qquad
\sigma_{f_{R0}} \equiv \frac{10^{-4.18}}{1.96}\simeq 3.37\times 10^{-5},
\label{eq:like_fR0}
\end{equation}
together with the consistency check that the predicted $(\mu_0,\Sigma_0)$ remain within the broad 68\% ranges
in Eq.~\eqref{eq:muSigma_bounds}. This yields a fully specified, non-empty inference layer that can be upgraded
to a full MCMC over LSS/CMB spectra once a dedicated solver is in place.

\paragraph{Status of the inference layer.}
The effective likelihood introduced here is intended as a conservative phenomenological layer,
not as a replacement for a full Boltzmann-level inference from CMB/LSS power spectra and
cross-correlations. Accordingly, the results in this section should be interpreted as translated late-time consistency constraints and parameter-structure diagnostics at the linear-response level.

\section{Results and discussion}
\label{sec:results}

\subsection{Translated bounds on the model parameters}
\label{subsec:translated_bounds}

For our action $f(R,T_\chi)=R+\alpha R^2+\xi R T_\chi+\beta T_\chi^2$, the background derivative with respect to $R$ reads
\begin{equation}
f_R = 1 + 2\alpha R + \xi T_\chi.
\end{equation}
Therefore the present-day deviation parameter is
\begin{equation}
f_{R0} \equiv f_R(R_0,T_{\chi 0})-1 = 2\alpha R_0 + \xi T_{\chi 0}.
\label{eq:fR0_model}
\end{equation}

Adopting a flat $\Lambda$CDM reference background with Planck-2018 parameters \cite{Planck2018Parameters},
the present-day Ricci scalar may be written as
\begin{equation}
R_0 = 6\left(2H_0^2+\dot H_0\right)
= 6\left(2-\frac{3}{2}\Omega_{m0}\right)H_0^2
\simeq 9.17\,H_0^2,
\label{eq:R0_LCDM}
\end{equation}
where we used $\dot H_0=-(3/2)\Omega_{m0}H_0^2$ for matter+$\Lambda$.

Using the benchmark constraint \eqref{eq:fR0_bound}, Eq.~\eqref{eq:fR0_model}
implies a stringent cap on the \emph{parameter combination} $2\alpha R_0+\xi T_{\chi0}$.
Two informative limiting cases are:

\paragraph{(i) Pure curvature correction at late times ($\xi=0$).}
Then $|f_{R0}|=|2\alpha R_0|$ and
\begin{equation}
|\alpha|H_0^2 < \frac{10^{-4.18}}{2(R_0/H_0^2)}
= \frac{10^{-4.18}}{2\times 9.17}
\simeq 3.60\times 10^{-6}.
\label{eq:alpha_bound}
\end{equation}

\paragraph{(ii) Pure selective coupling contribution at late times ($\alpha=0$).}
For a pressureless dark sector at $z\simeq 0$, $T_{\chi0}\simeq -\rho_{\chi0}$,
with $\rho_{\chi0}=\Omega_{c0}\rho_{c0}$ and $\rho_{c0}=3H_0^2/\kappa^2$.
Defining the dimensionless coupling $\tilde\xi \equiv \xi\rho_{c0}$,
Eq.~\eqref{eq:fR0_model} yields $|f_{R0}|\simeq |\tilde\xi|\Omega_{c0}$, so that
\begin{equation}
|\tilde\xi| < \frac{10^{-4.18}}{\Omega_{c0}}
\simeq 2.5\times 10^{-4},
\label{eq:xi_bound}
\end{equation}
where we used $\Omega_{c0}\simeq 0.264$ for the Planck-2018 baseline \cite{Planck2018Parameters}.

To assess how much observational leverage remains within the viable region implied by
Eq.~\eqref{eq:xi_bound}, we map the inter-benchmark spread of the reconstructed QS growth proxy
across the \((k,z)\) plane. This highlights the scale--redshift domain where growth measurements
can most efficiently discriminate among viable selective-coupling benchmarks.

\begin{figure}[!t]
  \centering
  \includegraphics[width=\linewidth]{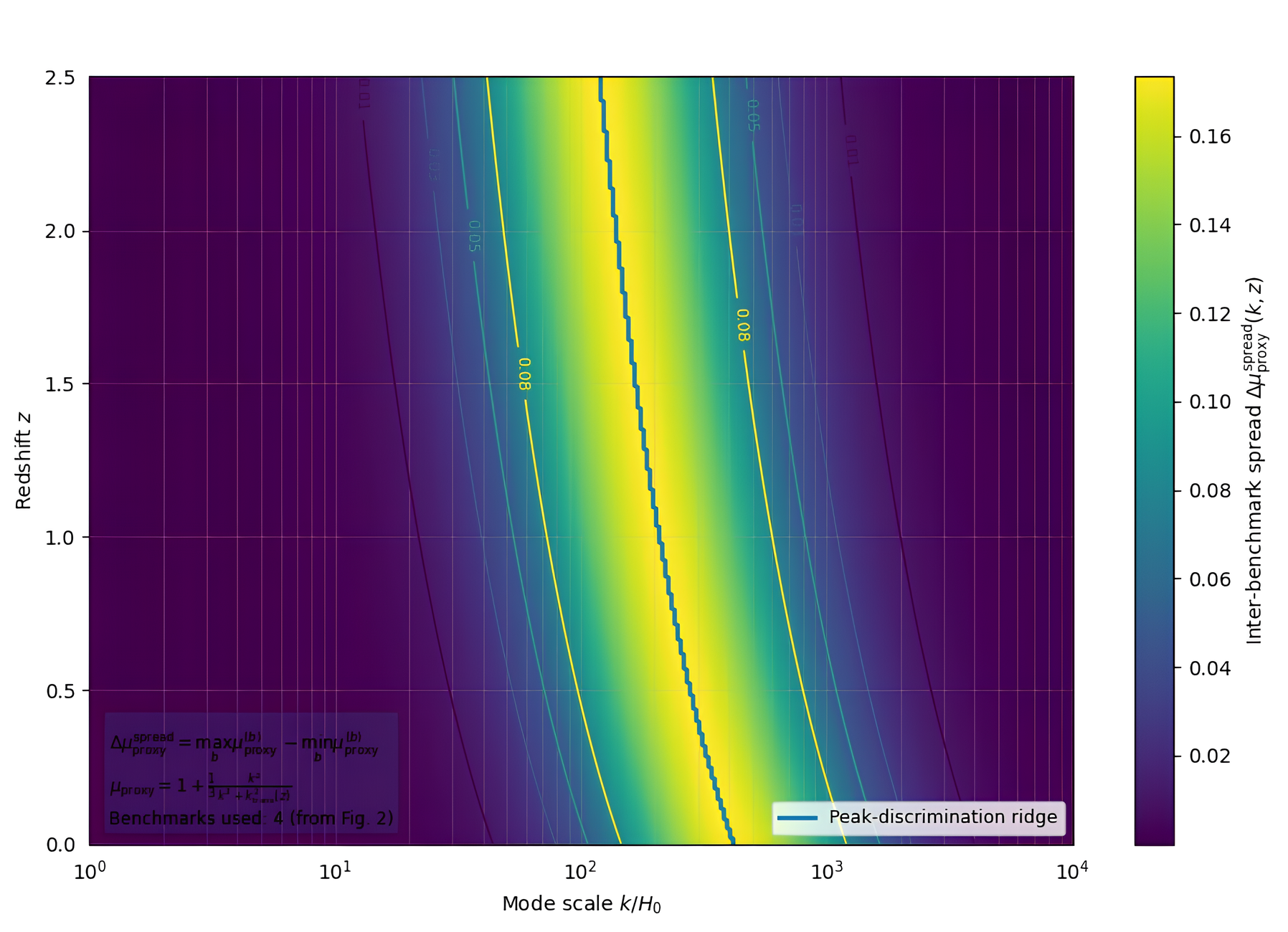}
  \caption{\textbf{Inter-benchmark discriminating-power map in the QS growth sector.}
  Two-dimensional map of the inter-benchmark spread of the reconstructed QS growth proxy,
  \(\Delta\mu_{\mathrm{proxy}}^{\mathrm{spread}}(k,z)\equiv
  \max_{b}\mu_{\mathrm{proxy}}^{(b)}(k,z)-\min_{b}\mu_{\mathrm{proxy}}^{(b)}(k,z)\),
  evaluated over the viable selective-coupling benchmark set \(b\) (four representative points, as in Fig.~2).
  The color scale shows where the QS growth response differs most strongly across viable benchmarks,
  providing a direct measure of \emph{discriminating power} in the \((k,z)\) plane.
  The thick curve marks the \emph{peak-discrimination ridge}, i.e.\ the locus where
  \(\Delta\mu_{\mathrm{proxy}}^{\mathrm{spread}}\) attains its maximum at fixed redshift.
  Overall, the figure isolates the mode-scale band and redshift range in which growth data are expected
  to be most sensitive to selective-coupling realizations while remaining consistent with the late-time bound
  in Eq.~\eqref{eq:xi_bound}.}
  \label{fig:6}
\end{figure}
\FloatBarrier

\noindent Figure~\ref{fig:6} therefore identifies the \((k,z)\) domain
where viable selective-coupling benchmarks are maximally separable through QS growth information.

\subsection{Consistency with $\mu_0$--$\Sigma_0$ bounds}
\label{subsec:consistency_muSigma}

Within the quasi-static linear regime, small $|f_{R0}|\ll 1$ implies that
departures in the effective clustering and lensing strengths scale as $\mathcal{O}(f_{R0})$
(up to the usual scale-dependent scalaron corrections) \cite{{DeFeliceTsujikawa2010fRReview}}.
Given $|f_{R0}| \lesssim 10^{-4.18}$, the predicted late-time $(\mu_0,\Sigma_0)$ remain automatically compatible
with the broader phenomenological constraints in Eq.~\eqref{eq:muSigma_bounds} \cite{Andrade2024MuSigma}.

\subsection{Interpretation for the dark sector}
\label{subsec:interpretation}

The translated bounds \eqref{eq:alpha_bound}--\eqref{eq:xi_bound} show that any viable late-time realization of the
present $f(R,T_\chi)$ model must be close to GR in terms of the scalar response $f_R$ today.
This does \emph{not} eliminate the model’s novelty: the selective coupling terms ($\xi RT_\chi$, $\beta T_\chi^2$)
can still leave distinctive fingerprints in (i) transition dynamics if the dark-sector equation of state departs from dust,
(ii) scale-dependent growth around $k/a\sim M_{\rm eff}(a)$the framework implies, and (iii) cross-correlations
where lensing and clustering respond differently to dark-sector stress. A dedicated numerical implementation will enable testing these
signatures beyond the present translated-linear bounds.

\section{Conclusions}
\label{sec:conclusions}

We developed a dark-sector selective extension of modified gravity in which the gravitational action depends on
the Ricci scalar $R$ and exclusively on the \emph{dark} trace $T_\chi$, $f(R,T_\chi)$, thereby avoiding direct visible-sector trace coupling in the fundamental action and reducing
the associated ambiguities and phenomenological tensions that arise in generic trace-coupled constructions. We derived the background
equations and linear perturbation relations in Newtonian gauge, including a closed elimination of $\delta R$
in favor of $\delta f_R$ and an explicit expression for $\delta f_T$ that controls the dark-sector selective
source entering the metric constraint. A key outcome is a gauge-consistent perturbation system for
$(\Phi,\Psi,\delta\chi,\delta f_R)$, from which the effective growth and lensing responses
$\mu(k,a)$ and $\Sigma(k,a)$ can be reconstructed in the quasi-static sub-horizon regime.

To obtain explicit late-time implications without requiring a dedicated Boltzmann-level likelihood pipeline, we adopted a conservative \emph{constraint-translation} strategy. Using current phenomenological bounds on $(\mu_0,\Sigma_0)$ together with
a representative 95\% limit on the present-day scalar response $|f_{R0}|$ from galaxy-clustering $\times$ CMB-lensing
cross-correlations, we translated these limits into constraints on our minimal polynomial realization
$f(R,T_\chi)=R+\alpha R^2+\xi R T_\chi+\beta T_\chi^2$. This yields \emph{translated} upper bounds on the late-time
parameter combination $2\alpha R_0+\xi T_{\chi0}$ and, in informative limits, on $|\alpha|H_0^2$ and the dimensionless
selective coupling $\tilde\xi\equiv \xi\rho_{c0}$ under the dust-level approximation $T_{\chi0}\simeq -\rho_{\chi0}$.
These translated bounds indicate that viable late-time realizations must remain close to GR in terms of the
present-day scalar response $f_R$, while still allowing qualitatively distinctive signatures through the selective
trace channel.

The principal physical novelty of the framework is that it separates the origin of deviations from GR into two
structurally distinct contributions in the quasi-static metric constraint: scalaron-mediated effects (curvature sector)
and dark-sector selective sourcing. This implies that combined growth and lensing datasets can, in principle, discriminate
a purely $f(R)$-like modification from an intrinsically dark-sector-driven departure. In particular, once a full linear
solver is implemented, the model predicts correlated but non-identical scale-dependent responses in clustering and lensing,
controlled by the transition scale $k/a\sim M_{\rm eff}(a)$ and by the selective source induced by $(\xi,\beta)$ through
$\delta f_T$ and $\Pi_{\rm DS}$.

A full-likelihood confrontation with CMB/LSS power spectra and cross-correlations requires a dedicated Boltzmann implementation and is beyond the scope of the present work. The results reported here are therefore restricted to translated late-time constraints in the linear/QS regime. Within this scope, the action-level formulation, exact exchange structure, and perturbation-level closure derived in this work provide a consistent basis for direct observational testing of dark-sector selective trace coupling.
\appendix

\section{Technical details and complementary derivations}
\label{app:technical}

This Appendix collects technical steps that are useful for reproducibility and for cross-checking the perturbation algebra,
and can be expanded if required by referees.

\subsection{Variation identities and conventions}
\label{app:variations}

We summarize the sign conventions, metric signature, and variation identities used throughout.
We work with signature $(-,+,+,+)$ and define
\begin{equation}
\Box \equiv g^{\mu\nu}\nabla_\mu\nabla_\nu,
\qquad
R_{\mu\nu} = \partial_\lambda \Gamma^\lambda{}_{\mu\nu} - \partial_\nu \Gamma^\lambda{}_{\mu\lambda}
+\Gamma^\lambda{}_{\lambda\rho}\Gamma^\rho{}_{\mu\nu}-\Gamma^\lambda{}_{\nu\rho}\Gamma^\rho{}_{\mu\lambda}.
\end{equation}
The variation of the Ricci scalar is written in the standard form
\begin{equation}
\delta R = R_{\mu\nu}\delta g^{\mu\nu} + g^{\mu\nu}\delta R_{\mu\nu},
\qquad
\delta R_{\mu\nu}=\nabla_\lambda \delta\Gamma^\lambda{}_{\mu\nu}-\nabla_\nu \delta\Gamma^\lambda{}_{\mu\lambda}.
\end{equation}

\subsection{Derivation of the anisotropic stress relation}
\label{app:slip_derivation}

We provide the intermediate steps leading to the slip relation
$\Phi-\Psi=\delta f_R/\bar f_R$ in Newtonian gauge by extracting the traceless spatial part of the
linearized field equations and using the absence of intrinsic anisotropic stress for the canonical dark scalar.

\subsection{Elimination of $\delta R$ in favor of $\delta f_R$}
\label{app:deltaR_elim}

Starting from $\delta f_R=f_{RR}\delta R+f_{RT}\delta T_\chi$ and assuming $f_{RR}\neq 0$, we derive
$\delta R=(\delta f_R-f_{RT}\delta T_\chi)/f_{RR}$ and insert it into $\delta f_T$ to obtain
Eq.~\eqref{eq:deltafT_closed}. We also discuss the special case $f_{RR}=0$, in which $\delta R is$ not
eliminated in the same way and the system must be reformulated.

\subsection{Quasi-static reduction and reconstruction formulae}
\label{app:QS_reduction}

We detail the steps used to obtain the QS relations for the potentials and for $\delta f_R$, including the
dominant-gradient approximation $\delta(\Box f_R)\simeq -(k^2/a^2)\delta f_R$, and we provide a checklist of
conditions under which the QS reduction is quantitatively reliable for a given dataset.

\subsection{Translated-constraint mapping details}
\label{app:constraint_translation}

We give the explicit mapping between the observational benchmark bound on $f_{R0}$ and the parameter combination
$2\alpha R_0+\xi T_{\chi0}$ in Eq.~\eqref{eq:fR0_model}, including the $\Lambda$CDM estimate for $R_0$ in
Eq.~\eqref{eq:R0_LCDM} and the dust approximation $T_{\chi0}\simeq -\rho_{\chi0}$. We also discuss how the mapping
changes if the dark sector departs from dust or if $T_\chi$ includes additional dark species.

\end{document}